\documentclass[journal,final,twoside,a4paper]{IEEEtran}
\usepackage{array}
\usepackage{cite}
\usepackage{amssymb}
\usepackage{amsmath}
\usepackage{amsthm}
\usepackage[dvips]{graphicx}
\usepackage{psfrag}
\usepackage{url}
\newcommand{\ignore}[1]{}

\newtheorem{proposition}{Proposition}
\newtheorem{lemma}{Lemma}

\usepackage{algorithm}
\usepackage{algorithmic}
\usepackage{subfigure}
\graphicspath{{./}{./Figures_eps/}}
\usepackage{enumerate}
\floatname{algorithm}{Algorithm}

\newcommand{\INDSTATE}[1][1]{\STATE\hspace{#1\algorithmicindent}}
\newcommand{\revN}[1]{{#1}}
\newcommand{\revNN}[1]{{#1}}
\newcommand{\rev}[1]{{#1}}

\usepackage{color}
\usepackage{textcomp}

\title{Optimized Distributed Inter-cell Interference
Coordination~(ICIC) Scheme using Projected Subgradient and
Network Flow Optimization}

\author{Akram Bin Sediq, Rainer Schoenen,
Halim Yanikomeroglu, and Gamini Senarath\\
\thanks{A. Bin Sediq, R. Schoenen, and H. Yanikomeroglu are with the Department of Systems and Computer Engineering, Carleton University, Ottawa, ON, Canada (e-mail: \{akram, rs, halim\}@sce.carleton.ca).}
\thanks{G. Senarath is with Huawei Technologies Canada Co., Ltd. (e-mail: gamini.senarath@huawei.com).}
\thanks{This work is supported in part by Huawei Canada Co., Ltd., in part by the Ontario Ministry of Economic Development and Innovation's ORF-RE (Ontario Research Fund - Research Excellence) program, and in part by an Ontario Graduate Scholarship in Science and Technology (OGSST). A preliminary version of this work was presented at IEEE PIMRC 2011~\cite{ABS_PIMRC2011}.}}

\newcounter{myenumi}

\allowdisplaybreaks

\begin{document}
\hfuzz=2pt
\vfuzz=2pt
  \mathchardef\mathcomma\mathcode`\,
  \mathcode`\,="8000

{\catcode`,=\active
  \gdef,{\mathcomma\discretionary{}{}{}}
}
\maketitle
\begin{abstract}
\label{abstract}
%
\revN{
In this paper, we tackle the problem of multi-cell resource scheduling, where the objective is to maximize the weighted sum-rate through inter-cell interference coordination (ICIC). The blanking method is used to mitigate the inter-cell interference, where a resource is either used with a predetermined transmit power or not used at all, i.e., blanked. This problem is known to be strongly NP-hard, which means that it is not only hard to solve in polynomial time, but it is also hard to find an approximation algorithm with guaranteed optimality gap. In this work, we identify special scenarios where a polynomial-time algorithm can be constructed to solve this problem with theoretical guarantees. In particular, we define a dominant interference environment, in which for each user the received power from each interferer is significantly greater than the aggregate received power from all other weaker interferers.

We show that the originally strongly NP-hard problem can be tightly relaxed to a linear programming problem in a dominant interference environment. Consequently, we propose a polynomial time distributed algorithm that is not only guaranteed to be tight in a dominant interference environment, but which also computes an upper bound on the optimality gap without additional computational complexity. The proposed scheme is based on the primal-decomposition method, where the problem is divided into a master-problem and multiple subproblems. We solve the master-problem iteratively using the projected-subgradient method. We also show that each subproblem has a special network flow structure. By exploiting this network structure, each subproblem is solved using the network-based optimization methods, which significantly reduce the complexity in comparison to the general-purpose convex or linear optimization methods. In comparison with baseline schemes, simulation results of the International Mobile Telecommunications-Advanced~(IMT-Advanced) scenarios show that the proposed scheme achieves higher gains in aggregate throughput, cell-edge throughput, and outage probability.
}

\end{abstract}


\section{Introduction}
\label{sec:introduction}
\ignore{Orthogonal Frequency Division Multiple Access (OFDMA) is the multiple access scheme for the 4th generation (4G) wireless cellular networks, namely, Long-Term Evolution (LTE), LTE-Advanced, and IEEE 802.16m. Among the advantages offered by OFDMA is its scheduling flexibility, since users can be scheduled in both time and frequency, which can be exploited to gain time, frequency, and multi-user diversity.}

Aggressive reuse is inevitable in cellular networks due to the scarcity of the radio resources. Reuse-1, in which all radio resources are reused in every sector, is an example of such an aggressive frequency reuse scheme. While reuse-1 can potentially achieve high aggregate system throughput, it jeopardizes the throughput experienced by users close to the cell\footnote{The terms ``cell'' and ``sector'' are used interchangeably in this work. }-edge, due to the excessive interference experienced by these users. Therefore, it is vital for the network to use robust and efficient interference mitigation techniques.


Conventionally, interference is mitigated by static resource partitioning and frequency/sector-planning, where close-by sectors are assigned orthogonal resources (clustering). A common example is reuse-3 (cluster size = 3 sectors)~\cite[Section 2.5]{Rappaport}, where adjacent sectors are assigned orthogonal channels, i.e., there is no inter-cell interference between adjacent sectors. Although such techniques can reduce inter-cell interference and improve cell-edge user throughput, they suffer from two major drawbacks. First of all, the aggregate network throughput is significantly reduced since each sector has only a fraction of the available resources, which is equal to the reciprocal of the reuse factor. Secondly, conventional frequency/sector-planning may not be possible in emerging wireless networks where new multi-tier network elements (such as relays, femto-/pico-base-stations, distributed antenna ports) are expected to be installed without prior planning in self-organizing networks~(SON)~\cite{Rahman2010}.

In order to reduce the effect of the first drawback, fractional frequency reuse (FFR) schemes have been proposed. The key idea in FFR is to assign lower reuse factor for users near the cell-center and higher reuse factor for users at the cell-edge. The motivation behind such a scheme is that cell-edge users are more vulnerable to inter-cell interference than cell-center users. Soft Frequency Reuse (SFR)~\cite{SFR2005} and Partial Frequency Reuse (PFR)~\cite{Sternad2003} are two examples of FFR. A comparative study between reuse-1, reuse-3, PFR, and SFR is provided in \cite{ComparingPFRSFR,VTC2007Comparison}. While FFR schemes recover some of the throughput loss due to partitioning, they require frequency/sector planning a priori, which is not desirable in future cellular networks as mentioned earlier. As a result, developing efficient\ignore{and robust} dynamic inter-cell interference coordination (ICIC) schemes is vital to the success of future cellular networks~\cite{Rahman2010}.

One approach to tackle the ICIC problem is to devise adaptive FFR or SFR schemes. In~\cite{AdaptiveFFR}, an adaptive FFR scheme is proposed where each base-station~(BS) chooses one of several reuse modes. A dynamic and centralized\footnote{\revN{\label{difference_cent_dist}In this paper, an algorithm is categorized as a centralized algorithm if it requires a centralized entity to coordinate or to perform either partial or full execution of the algorithm. In contrast, a distributed algorithm is one that can be executed in each sector with limited message exchange between the sectors, without the requirement of having a centralized entity.}} FFR scheme is proposed in~\cite{Hussain2009} that outperforms conventional FFR schemes in terms of the total system throughput. In~\cite{Zhang2008}, the authors propose softer frequency reuse, which is a heuristic algorithm based on modifying the proportional fair algorithm and the SFR scheme. In~\cite{Mao2008}, a heuristic algorithm based on adaptive SFR is proposed for the uplink. In~\cite{Stoylar2009}, the authors propose gradient-based distributed schemes that create SFR patterns, in an effort to achieve local maximization of the network utility. \revN{To enable gradient computation, 
the rate adaptation function that maps signal-to-interference-and-noise-ratio~(SINR) to rate is required to be differentiable. However, this requirement is not currently feasible since rate adaption is performed in current cellular standards by using adaptive modulation and coding~(AMC) with a discrete set of modulation and coding schemes~\cite{DBNRS_PIMRC10}, which results in a non-differentiable rate adaptation function.}

Another approach to tackle the ICIC problem is to remove the limitations imposed by the FFR schemes and view the ICIC problem as a multi-cell scheduling problem. In~\cite{Li2006}, the authors propose a suboptimal \revN{centralized} algorithm where a radio network controller is assumed to be connected to all BSs. The authors conclude that the problem is NP-hard and they resort to heuristics that show improvement in the performance. In~\cite{Chang2009}, a graph-theoretic approach is taken to develop an ICIC scheme in which information about the interference experienced by each user terminal~(UT) is inferred from the diversity set of that UT. In~\cite{Ellenbeck2008}, a game-theoretic approach is pursued and \revN{an autonomous} decentralized algorithm is developed. The proposed algorithm converges to a Nash equilibrium in a simplified cellular system. Nevertheless, a significant gap is observed between the proposed algorithm and the globally optimum one, which is computationally demanding. In~\cite{Rahman2010}, a partly distributed two-level ICIC scheme is proposed where a centralized entity is required to solve a binary linear optimization problem, which is generally not solvable in polynomial time.

\revN{The focus of this paper is to overcome the drawbacks of the existing schemes mentioned above by developing a distributed ICIC scheme, that can work with any AMC scheme, and yet provide theoretical guarantee}. {\label{balnking_intro} \revN{In particular, we tackle the problem of distributed multi-cell resource scheduling, where the objective is to maximize the weighted sum-rate. The blanking method is used to mitigate the inter-cell interference, where a resource is either used with a predetermined transmit power or not used at all, i.e., blanked, similar to~\cite{Rahman2010}. This problem is known to be strongly NP-hard, which means that it is not only hard to solve in polynomial time, but it is also hard to find an approximation algorithm with guaranteed optimality gap~\cite{TomLuo2008}.}} The main contributions can be summarized as follows:
\label{contributions}
\revN{
    \begin{enumerate}
    \item We show that the originally strongly NP-hard problem can be tightly relaxed to a linear programming problem in a dominant interference environment, in which for each user the received power from each interferer is significantly greater than the aggregate received power from all other weaker interferers\footnote{The definition of a dominant interference environment is illustrated by the following example. Assume there are four interferers, and let $Pr^{(k)}$ denote the received power from the $k^{\textrm{th}}$ interferer. Assume further that $Pr^{(1)}>Pr^{(2)}>Pr^{(3)}>Pr^{(4)}$. Such an environment is called a dominant interference environment if and only if the following conditions are satisfied: $Pr^{(1)}>>Pr^{(2)}+Pr^{(3)}+Pr^{(4)}$, $Pr^{(2)}>>Pr^{(3)}+Pr^{(4)}$, and $Pr^{(3)}>>Pr^{(4)}$.}. The tightness of the relaxation is shown by proving that the percentage of optimal relaxed variables that assume binary values is bounded below by 
    $\frac{\tilde{K}(\bar{M}-1)}{(\tilde{K} + 1)\bar{M}+1} \cdot 100 \%$, where $\bar{M}$ is the average number of UTs per sector and $\tilde{K}$ is the number of neighboring interferers. Consequently, we devise a polynomial-time distributed algorithm that is not only guaranteed to be tight, but which also computes an upper bound on the optimality gap without additional computational complexity. The proposed algorithm does not require a central controller and it can be used with any AMC scheme, including AMC schemes that result in non-differentiable discrete rate adaption functions.
    \item We demonstrate that considered optimization problem for a dominant interference environment can be transformed into an equivalent minimum-cost network flow~(MCNF) optimization problem. Thus, we can use network-based algorithms which have significantly reduced complexity as compared to the general-purpose convex or linear optimization algorithms~\cite[p. 402]{Network_flows}.\revN{\label{MCNF_reference} While MCNF optimization tools have been used in the literature to solve resource allocation problem in single-cell networks, e.g.,\cite{Tao2013_MCNF} and~\cite{Zaki2011_MCNF}, as far as we know, our work is the first in literature to use these optimization tools in ICIC, i.e., in multi-cell resource allocation problems.}
    \end{enumerate}
    }

\section{System Model}
\label{sec:SystemModel}
\begin{figure}[!ht]
    \centering
    \resizebox{.46\textwidth}{!}{\includegraphics{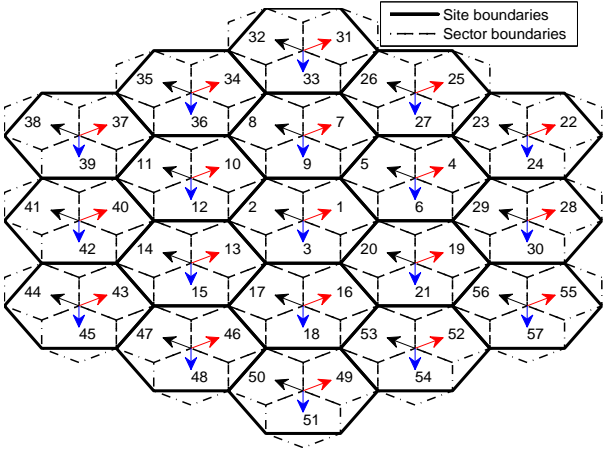}}
    \caption{The considered network layout which consists of 19 cell sites and 3 sectors per site.}
    \label{Fig:Layout}
    \end{figure}

We consider the network model described by the International Mobile Telecommunications-Advanced~(IMT-Advanced) evaluation guidelines~\cite{IMTAdvanced}. Based on these guidelines, the considered network consists of $K$ sectors served by $K/3$ BSs as shown in Fig.~\ref{Fig:Layout}. Each BS is equipped with a tri-sector antenna to serve a cell-site that consists of 3 sectors. Each BS can communicate with its neighboring BSs; this is supported in most cellular network standards, e.g., using the R8 interface in IEEE~802.16m
standard and the X2 interface in Long-Term Evolution (LTE) and LTE-advanced standards. Hexagonal sectors are considered herein according to IMT-Advanced guidelines; nevertheless, the proposed scheme works also for arbitrary sector shapes.  We focus on the downlink scenario in this paper. For convenience, the symbols used in this section and onward are provided in Table~\ref{table_symbols}.

 \begin{table}[!ht]
 	 \centering
 	 \caption{List of symbols}
  		\label{table_symbols}
\scalebox{0.9}{
 	 \begin{tabular}{l l | l}
 \hline 
 	& $M^{(k)}$ & Number of UTs in sector $k$\\
 	& $N$ & Number of available RBs\\
 	& $K$ & Number of sectors\\
    &$\mathcal{M}^{(k)}$ & Set of indices of users in sector $k$, i.e., $\mathcal{M}^{(k)}\triangleq\{1,\ldots,M^{(k)}\}$ \\
    &$\mathcal{N}$ & Set of indices of available RBs, i.e., $\mathcal{N}\triangleq\{1,\ldots,N\}$ \\
    &$\mathcal{K}$ & Set of indices of all sectors in the network, i.e., $\mathcal{K}\triangleq\{1,\ldots,K\}$ \\
    & $\tilde{\cal K}^{(k)}$ & Set of indices of the first-tier interfering sectors seen by sector $k$\\
    & $\tilde{K}$ & The cardinality of $\tilde{\cal K}^{(k)}$, which is assumed to be the same $\forall k$\\
 	& $m$ & UT index; $m \in \mathcal{M}^{(k)}$\\
 	& $n$ & RB index; $n \in \mathcal{N}$ \\
 	& $k$ & Sector index; $k \in \mathcal{K}$ \\
    & $H_{m,n}^{(k,\tilde k)}$ &  Channel gain from sector $\tilde k$ on RB $n$ to UT $m$ served by sector $k$\\
 	& ${R}_{m,n}^{(k)}$ & Achievable rate on RB $n$ for UT $m$ in sector $k$ \\
 	& $r_{m,n}^{(k)}$ & Achievable rate on RB $n$ for UT $m$ in sector $k$, if RB $n$ is reused\\
    & & by all other interfering sectors\\
 	& ${\tilde r}_{m,n}^{(k,\tilde k)}$ & Additional rate on RB $n$ for UT $m$ in sector $k$, gained if RB $n$ \\
    & & is \revN{blanked} in sector $\tilde k$ \\
 	& ${\bar R}_{m}^{(k)}$ & Average rate for UT $m$ in sector $k$ \\
 	& $\Gamma_{m,n}^{(k)}$ & SINR on RB $n$ of UT $m$ in sector $k$ \\
 	& $\gamma_{m,n}^{(k)}$ & SINR on RB $n$ of UT $m$ in sector $k$, if all sectors reuse RB $n$ \\
 	& $\tilde \gamma_{m,n}^{(k,\tilde k)}$ & SINR on RB $n$ of UT $m$ in sector $k$, if sector $\tilde k$ \revN{blanks} RB $n$ \\
 	& $x_{m,n}^{(k)}$ & Binary variable indicating whether RB $n$ is assigned to UT $m$ in \\
    & & sector $k$ or not\\
 	& $I_n^{(k)}$ & Binary variable indicating whether RB $n$ is \revN{blanked} in sector $k$ \\
     & & or not\\
 		\hline
 	 \end{tabular}
 }
 	\end{table}

Orthogonal Frequency Division Multiple Access (OFDMA) is used as the multiple access scheme, since it is 
adopted in most of the contemporary cellular standards.
 Among the advantages offered by OFDMA is its scheduling flexibility, since users can be scheduled in both time and frequency, which can be exploited to gain time, frequency, and multi-user diversity.
The time and frequency radio resources are grouped into time-frequency resource
blocks (RBs). RB is the smallest radio resource block that can be
scheduled to a UT 
and it consists of $N_s$
OFDM symbols in the time dimension and $N_f$ sub-carriers in the frequency dimension\ignore{ (in LTE, $N_s=7$ and $N_f=12$)}. The total number of RBs is denoted by $N$. The number of UTs in sector $k$ is denoted by $M^{(k)}$. Both the BSs and the UTs are assumed to have single antenna each. Similar to~\cite{Rahman2010}, we assume that each UT estimates and reports to its serving BS the channel from its serving sector's antenna and from the first-tier interfering sectors. The SINR observed by UT $m\in\mathcal{M}^{(k)}\triangleq\{1,\ldots,M^{(k)}\}$ in sector $k\in\mathcal{K}\triangleq\{1,\ldots,K\}$ on RB $n\in\mathcal{N}\triangleq\{1,\ldots,N\}$  is given by~\cite{Rahman2010} 
    \begin{equation}
    \label{SINR}
     \Gamma_{m,n}^{(k)} = \frac{{{P_C}H_{m,n}^{(k,k)}}}{{{P_C}\sum\limits_{\tilde k=1, \tilde k \neq k}^K {\Bigl( {1 - I_n^{(\tilde k)}} \Bigr)H_{m,n}^{(k,\tilde k)}}  + {P_N}}},
    \end{equation}
where {\label{P_C_answer}\revN{$I_n^{(k)}$ is a binary variable indicating whether RB $n$ is \revN{blanked}, i.e., not used, in sector $k$ ($I_n^{(k)}=1$) or not ($I_n^{(k)}=0$), $P_C$ represents the transmitted power per RB and it is assumed to be the same for all non-\revN{blanked} RBs}}, $P_N$ represents the thermal noise power per RB, and $H_{m,n}^{(k,\tilde k)}$ represents the channel gain from sector $\tilde k$ on RB $n$ to UT $m$ served by sector $k$.
Large scale channel variations (due to distance-dependant attenuation and shadowing), antenna gains, and multipath fading are all captured in $H_{m,n}^{(k,\tilde k)}$. The achievable rate on RB $n$ of UT $m$ in sector $k$ is given by
     \begin{equation}
    \label{R_mnk}
         R_{m,n}^{(k)} = f\bigl( {\Gamma_{m,n}^{(k)}} \bigr) \quad \textrm{bit/sec},
    \end{equation}
    where $f(\cdot)$ is the AMC \revN{rate adaptation} function that maps SINR to rate. The function $f(\cdot)$ is assumed to be nondecreasing with $f(0)=0$, possibly discontinuous, which is the case for all practical AMC schemes. 


\section{Problem Statement}
\label{sec:ProblemStatement}
Let us consider a generic scheduler implemented in sector $k$, without ICIC, such that it maximizes the weighted sum of the UTs' rates in sector $k$. This scheduler is implemented by solving the following optimization problem for each RB $n$ in every sub-frame $t$:
    \begin{subequations}
    \label{scheduler}
    \begin{align}
    \underset{\substack{\revN{\{x^{(k)}_{m,n}(t):  m \in \mathcal{M}^{(k)}\} }}}
    {\text{maximize}}& \quad \sum_{m=1}^{M^{(k)}} w^{(k)}_m(t) x^{(k)}_{m,n}(t)R^{(k)}_{m,n}(t)
    \label{scheduler:objective}\\
        \text{subject to}& \quad \sum_{m=1}^{M^{(k)}}x^{(k)}_{m,n}(t)= 1, \quad \label{scheduler:Constraint1}\\
        & \quad x^{(k)}_{m,n}(t)\in \{0,1\}, \quad \forall m\label{scheduler:Constraint2},
    \end{align}
    \end{subequations}
where $w^{(k)}_m(t)$ is the weight of UT $m$ in sector $k$, $\{x_{m,n}^{(k)}(t)\}$ are the binary decision variables such that $x_{m,n}^{(k)}(t)=1$ when RB $n$ is assigned to UT $m$ in sector $k$, and $x_{m,n}^{(k)}(t)=0$ otherwise. Constraint~\eqref{scheduler:Constraint1} ensures that each RB is assigned to only one user in sector $k$.

The scheduler described above can be used \revN{to} control the desired fairness-throughput tradeoff in the long-term average rates by updating the weights $\{w^{(k)}_m(t)\}$ in every sub-frame.
To elaborate, \revN{let $R^{(k)}_m(t)$ denote the data rate scheduled to user $m$ in sector $k$ at sub-frame $t$, i.e., $R^{(k)}_m(t)=\sum\limits_{n=1}^{N} x^{(k)}_{m,n}(t) R^{(k)}_{m,n}(t)$ and} let $\bar R^{(k)}_m(t)$ denote the average rate of user $m$ in sector $k$ in sub-frame $t$, averaged over a window of $t_c$ sub-frames using exponentially-weighted low-pass filter~\cite{Kushner}, i.e., 
    \revN{
    \begin{align}
    \label{R_avg_exponential}
      \bar R^{(k)}_m(t)&= \frac{1}{t_c}\sum\limits_{i=0}^{t} (1-\frac{1}{t_c})^{i-t}R^{(k)}_m(i)\nonumber\\ 
        &= \Bigl(1-\frac{1}{t_c}\Bigr)\bar R^{(k)}_m(t-1) + \frac{1}{t_c}R^{(k)}_m(t).
     \end{align}
     }
Moreover, let $\bar R^{(k)}_m$ denote the long-term (steady-state) average rate of user $m$ in sector $k$, i.e., $\bar R^{(k)}_m= \displaystyle\lim_{t \rightarrow \infty} \bar R^{(k)}_m(t)$. A common approach to achieve various tradeoffs between fairness and aggregate throughput is to maximize a concave utility $U(\bar R^{(k)}_1,\dots,\bar R^{(k)}_{M^{(k)}} )$ that incorporates both fairness and aggregate throughput. It is shown in~\cite{StoylarGradient2005} and~\cite{SongPart2} that such a maximization can be achieved by solving the instantaneous optimization problem~\eqref{scheduler} in every $t$ with the weights $\{w^{(k)}_m(t)\}$ chosen to be the marginal utility, i.e., $w^{(k)}_m(t)=\frac{\partial U (\bar R^{(k)}_1(t-1),\dots,\bar R^{(k)}_{M^{(k)}}(t-1) )}{\partial \bar R^{(k)}_m(t-1)}$. Optimality in the long-term average rates is attained for sufficiently large $t_c$ and under ergodic channels.
A widely used utility in this context is the $\alpha$-fair utility defined for $\alpha\geq 0$ as~\cite{GeneralizedPF}
    \begin{equation}
    \label{Ualpha}
    U_{\alpha} \bigl(\bar R^{(k)}_1,\dots,\bar R^{(k)}_{M^{(k)}} \bigr)=\left\{\begin{array}{ll}
        \sum_{m=1}^{M^{(k)}} \log \bar R^{(k)}_m,& \alpha=1,\\
        \frac{1}{1-\alpha}\sum_{m=1}^{M^{(k)}} (\bar R^{(k)}_m)^{1-\alpha},& \alpha \neq 1.
    \end{array}
\right.
    \end{equation}
Based on this choice of utility, the user weights are given by 
$w^{(k)}_m(t)=\bigl(\bar R^{(k)}_m(t-1)\bigr)^{-\alpha}$,
where various degrees of fairness-throughput tradeoff can be achieved by varying the fairness exponent, $\alpha$~\cite{GeneralizedPF}. In general, as $\alpha$ increases, the degree of fairness increases and the total throughput decreases, and vice versa (cf.~\cite{ChiangFairness}). For example, max-SINR scheduler corresponds to the case where $\alpha=0$, proportional-fair scheduler~(cf.~\cite{KellyPF,OpportunisticBeamforming}) corresponds to the case where $\alpha=1$, and max-min scheduler corresponds to the case where $\alpha \rightarrow \infty$. \revN{One can also use the recently proposed utility in~\cite{JTWC2013} that achieves the optimum tradeoff between total throughput and Jain's fairness index by setting the user weights as $w^{(k)}_m(t) = \beta-\bar R^{(k)}_m(t-1)$, where $\beta$ is a non-negative parameter, such that higher $\beta$ results in higher total throughput and less fairness, and vice versa (cf.~\cite{JTWC2013})}.

Another approach to vary the degree of fairness-throughput tradeoff is by imposing a minimum long-term average rate~\big($\bar R_{\min}$\big), varying~$\bar R_{\min}$~\cite{EffeciencyFairness3},     and updating $\{w^{(k)}_m(t)\}$ according to the algorithm explained in \cite{Stolyar_PF}. We investigate both approaches in this paper.

It can be shown that the solution to~\eqref{scheduler} is given by
    \begin{equation}
    \label{schedulerSolution}
    \begin{array}{l}
     x_{m,n}^{\star (k)}(t) = \left\{ {\begin{array}{*{20}{c}}
       {1,} & {m = \arg \mathop {\max }\limits_m {w^{(k)}_m(t) R_{m,n}^{(k)}}(t),} \\
       {0,} & {m \ne \arg \mathop {\max }\limits_m {w^{(k)}_m(t) R_{m,n}^{(k)}}(t).} \\
    \end{array}} \right. \\
     \end{array}
    \end{equation}
    To simplify the notation, we drop the sub-frame index $(t)$ in the rest of the paper.


    In order to improve the performance of the above mentioned scheduler, we seek an ICIC scheme that coordinates the scheduling in all sectors in order to maximize the weighted sum of the rates of all users in the network. The optimization problem for each RB $n$ to attain this goal can be formulated as
    \begin{subequations}
    \label{ICICNPHard}
    \begin{align}
    \underset{\substack{\revN{\{x^{(k)}_{m,n},I^{(k)}_{n} : }\\ \revN{k \in \mathcal{K}, m \in \mathcal{M}^{(k)}}\}}}
    {\text{maximize}}& \quad \sum_{k=1}^{K}\sum_{m=1}^{M^{(k)}} w^{(k)}_m x^{(k)}_{m,n} R^{(k)}_{m,n}
    \label{ICICNPHard:objective}\\
        \text{subject to}& \quad \sum_{m=1}^{M^{(k)}}x^{(k)}_{m,n}= 1-I^{(k)}_n, \quad \forall k, \label{ICICNPHard:Constraint3}\\
        & \quad x^{(k)}_{m,n}, I^{(k)}_n \in \{0,1\}, \qquad \forall k, m,~\label{ICICNPHard:Constraint4}
    \end{align}
    \end{subequations}
    where the binary variables $\{I_n^{(k)}\}$ are introduced such that $I_n^{(k)}=1$ when RB $n$ is \revN{blanked} in sector $k$ and $I_n^{(k)}=0$ otherwise. Constraint~\eqref{ICICNPHard:Constraint3} ensures that each RB $n$ is assigned to only one UT in sector $k$, given that RB $n$ is not \revN{blanked} in sector $k$.

The network-wide optimization problem in~\eqref{ICICNPHard} is difficult to solve for the following reasons. First of all, this problem belongs to the class of non-linear binary combinatorial optimization problems ($R^{(k)}_{m,n}$ is a nonlinear function of $I^{(k)}_n$, cf.~\eqref{SINR} and~\eqref{R_mnk}), which are generally difficult to solve in polynomial time. Indeed, \eqref{ICICNPHard} is strongly NP-hard even for the case of a single user per sector \cite{TomLuo2008}\footnote{{\revN{For instance, consider the special case explained  in~\cite{TomLuo2008}, where all cross-channel gains are either 0 or $\infty$ and there is one UT per sector; in this case, the optimum transmission powers are binary, yet it is strongly NP-hard to determine for each UT whether to transmit with full power or not to transmit at all.}}\label{stronlgy_np_hard}}. This means that not only~\eqref{ICICNPHard} is  hard to solve in polynomial time, but it is also hard to find an approximation algorithm with guaranteed optimality gap. Moreover, the objective function is dependant on the AMC \revN{rate adaptation} function that is used. \ignore{Hence, an optimal solution for a given AMC \revN{rate adaptation} function may not be optimal for another AMC strategy.} Since AMC \revN{rate adaptation} functions are operator dependant, it is desirable to develop an algorithm that is independent of the chosen AMC \revN{rate adaptation} function. Finally, it is desirable to solve \eqref{ICICNPHard} in a distributed manner since most contemporary standards (such as LTE, LTE-Advanced, IEEE 802.16m) do not support a central controller.
The main contribution of this work is to circumvent these difficulties and develop  an efficient algorithm that obtain a near-optimal solution of~\eqref{ICICNPHard}. 
The development of such an algorithm is explained in the following section.

\section{Proposed Algorithm}
\label{sec:ProposedAlgorithm}
In this section, we show the steps used to develop the proposed algorithm. We start in Section~\ref{sec:BoundOptimization} by introducing a bound and argue that this bound is a good metric for optimization, \revN{especially for dominant interference environment}. Using the bound, we get a binary non-linear optimization problem and we transform it into an equivalent binary linear optimization problem in Section~\ref{sec:BNLP2BLP}.
Next, we relax the resulted binary linear optimization problem into a linear optimization problem in Section~\ref{sec:Relaxation}. Then, we devise a distributed algorithm in Section~\ref{sec:decomposition} by decomposing the problem into a master problem and multiple subproblems using the primal-decomposition method. To reduce the computational complexity, 
we show in Section~\ref{sec:MCNF} that each subproblem is amenable to powerful network flow optimization methods. We then present a pseudocode of the proposed algorithm in Section~\ref{sec:Pseudo} and discuss its complexity in Section~\ref{sec:complexity}. \ignore{Finally, we discuss the applicability of our scheme in heterogenous networks in Section~\ref{sec:eICIC}.}\revN{Finally, we highlight the advantages of the distributed implementation of the proposed scheme as compared to a centralized implementation of the same scheme in Section~\ref{advantages_distributed}}. \ignore{A block diagram that shows the roadmap of this section is shown in Fig.~\ref{Block diagram}.
    \begin{figure}[!htt]
    \centering
    \psfrag{a}[][][1.]{Difficult problem \eqref{ICICNPHard}}
    \psfrag{b}[][][1.]{(strongly NP-hard)}
    \psfrag{c}[][][1]{Rate bound \eqref{Rate_bound}}
    \psfrag{d}[][][1]{}
    \psfrag{e}[][][1]{Binary non-linear}
    \psfrag{f}[][][1]{optimization \eqref{ICICINLP}}
    \psfrag{g}[][][1]{Binary linear}
    \psfrag{h}[][][1]{optimization \eqref{ICICILP}}
    \psfrag{i}[][][1]{Relaxed linear}
    \psfrag{j}[][][1]{optimization}
    \psfrag{k}[][][1]{Primal decomposition}
    \psfrag{l}[][][1]{Master problem~\eqref{ICICMaster}}
    \psfrag{m}[][][1]{Subgradient method}
\psfrag{n}[][][1]{Subproblem~\eqref{ICICMCNF}}

    \psfrag{o}[][][1]{Minimum cost flow}
    \psfrag{p}[][][1]{Proposed algorithm}
    \psfrag{q}[][][0.92]{Pseudocode in Sec.~\ref{sec:Pseudo}}
    \psfrag{r}[][][1]{Checking the approx.}
    \psfrag{s}[][][1]{Tables~\ref{table:OptimalityGap}.a and~\ref{table:OptimalityGap}.b}
    \includegraphics[scale=0.75]{roadmap9.eps}\
    \caption{Block diagram which shows the steps used to develop the proposed algorithm.}
    \label{Block diagram}
    \end{figure}
    }
\subsection{Bound Optimization}
\label{sec:BoundOptimization}
     The SINR expression in~\eqref{SINR} can be lower-bounded as
    \begin{equation}
    \label{SINR_bound}
    \begin{array}{l}
     \Gamma_{m,n}^{(k)} \ge \revN{\tilde \Gamma_{m,n}^{(k)}} = \frac{{{P_C}H_{m,n}^{(k,k)}}}{{{P_C}\sum\limits_
     { \tilde k = 1,\tilde k \neq k}^K {H_{m,n}^{(k,\tilde k)}}  - \mathop {\max }\limits_{\tilde k \in \tilde{\cal K}^{(k)}} I_n^{(\tilde k)}{P_C}H_{m,n}^{(k,\tilde k)} + {P_N}}},
     \end{array}
    \end{equation}
    where $\tilde{\cal K}^{(k)}$ is the set of indices of the neighboring sectors of sector $k$, which are considered in this work to be
    the first-tier interfering sectors seen by sector $k$; e.g., in Fig. \ref{Fig:Layout}, ${\cal K}^{(3)}=\{1,2,13,17,16,20\}$\footnote{We assume a wraparound layout so each sector has 6 first-tier interfering sectors; e.g., ${\cal K}^{(25)}=\{26,27,23,51,50,48\}$.}. \revN{The cardinality of $\tilde{\cal K}^{(k)}$ is assumed to be the same for all sectors and it is denoted by $\tilde K$, i.e., $|\tilde{\cal K}^{(k)}| = \tilde K, \forall k$.}
\rev{The bound in \eqref{SINR_bound} is obtained by assuming that RB $n$ is used in all sectors, except at most one sector. When \revN{RB $n$ is blanked in more than one interfering sector}, only the most dominant blanked interferer is considered}\footnote{Note that this bound considers the most dominant \emph{blanked} interferer which is not necessarily the most dominant interferer when the most dominant interferer \revN{does not blank RB $n$} (i.e., when it is not turned off). This differentiates~\eqref{SINR_bound} from the expression considered in~\cite{Li2006}, where the most dominant interferer is considered.}. This bound is exact if the number of \revN{blanked} interferers is less or equal to one and it is tight for small number of \revN{blanked} interferers. \revN{The bound is also tight for dominant interference environment, where the received power from each interferer to each user is significantly greater than the aggregate received power from all other weaker interferers.}

{\label{bound_advantages} The bound \revN{in \eqref{SINR_bound} is a good metric} to be used for optimization (maximization) for the following reasons. First of all, if the bound is increased by $\Delta$, then the exact expression given by~\eqref{SINR} will also increase by at least $\Delta$ (proof is given in Appendix~\ref{Appendix:Bound}). \rev{Moreover, it is already observed in the literature that~\revN{blanking a particular RB} in more than two sectors can degrade the overall system performance (cf.~\cite{Rahman2010}).} Finally, and most importantly, based on this bound, we can develop a distributed optimization framework that is applicable to a wide range of schedulers and AMC strategies for \revN{dominant interference environments}. This framework can be implemented very efficiently, and can achieve near-optimal performance, as we will see later.

\rev{Using the SINR bound given in~\eqref{SINR_bound}, we now construct a similar bound on the rates, $\{R_{m,n}^{(k)}\}$, given in~\eqref{R_mnk}. To do so, we define $\gamma_{m,n}^{(k)}$ and $r_{m,n}^{(k)}$ as the SINR and the corresponding achievable rate on RB $n$ of UT $m$ in sector $k$, if all sectors use RB $n$, i.e.,
         \begin{equation}
    \label{gamma1}
      \gamma_{m,n}^{(k)}={\frac{{{P_C}H_{m,n}^{k, k}}}{{{P_C}\sum\limits_{\tilde k \neq k} {H_{m,n}^{(k,\tilde k)}} + {P_N}}}},
      r_{m,n}^{(k)} = f\Big( \gamma_{m,n}^{(k)} \Big).
     \end{equation}
    Similarly, we define $\tilde \gamma_{m,n}^{(k,\tilde k)}$ and ${\tilde r}_{m,n}^{(k,\tilde k)}$ as the SINR and the {\label{additional_answer} \revN{additional}} rate on RB $n$ of UT $m$ in sector $k$, if only sector $\tilde k \in \tilde{\cal K}^{(k)}$ \revN{blanks} RB $n$, i.e.,
    \begin{equation}
    \label{gamma2}
   \begin{array}{l}
    \tilde \gamma_{m,n}^{(k,\tilde k)}={\frac{{{P_C}H_{m,n}^{(k,k)}}}{{{P_C}\sum\limits_{\hat k \neq k} {H_{m,n}^{k,\hat k}} - {P_C}H_{m,n}^{(k,\tilde k)} + {P_N}}}},\\
    \tilde r_{m,n}^{(k,\tilde k)} = f\Big( \tilde \gamma_{m,n}^{(k,\tilde k)} \Big) - r_{m,n}^{(k)}.
     \end{array}
     \end{equation}
{\label{Rate_bound_answer} \revN{Using the definitions of $\gamma_{m,n}^{(k)}$ and $\tilde \gamma_{m,n}^{(k,\tilde k)}$ given in~\eqref{gamma1} and~\eqref{gamma2}, respectively, the SINR bound given by~\eqref{SINR_bound} can be also written as
\begin{equation}
    \label{SINR_bound_2}
\Gamma_{m,n}^{(k)} \ge {\tilde \Gamma_{m,n}^{(k)}} = \max \Big(\gamma_{m,n}^{(k)}, \max_{\tilde k} I_n^{(\tilde k)} \tilde \gamma_{m,n}^{(k,\tilde k)} \Big),
\end{equation}
where the equivalence between \eqref{SINR_bound} and~\eqref{SINR_bound_2} stems from the fact that $\{I_n^{(\tilde k)}\}$ assume binary values.

    Using \eqref{gamma1} and \eqref{gamma2}, a bound on the rates, $\{R_{m,n}^{(k)}\}$,  given by \eqref{R_mnk} is constructed based on the SINR bound given by \eqref{SINR_bound_2} as
    \begin{equation}
    \label{Rate_bound}
     \begin{array}{ll}
R_{m,n}^{(k)} = f(\Gamma_{m,n}^{(k)}) &\ge f({\tilde \Gamma_{m,n}^{(k)}}) \\
&= f\Big(\max\Big({\gamma_{m,n}^{(k)}, \max_{\tilde k} I_n^{(\tilde k)} \tilde \gamma_{m,n}^{(k,\tilde k)}}\Big)\Big) \\
&= \max\Big({r_{m,n}^{(k)}, \max_{\tilde k} I_n^{(\tilde k)} (r_{m,n}^{(k)} + \tilde r_{m,n}^{(k,\tilde k)})}\Big) \\
&= r_{m,n}^{(k)} + \mathop {\max }\limits_{\tilde k \in \tilde{\cal K}^{(k)}} I_n^{(\tilde k)}{\tilde r}_{m,n}^{(k, \tilde k)}, 
\end{array}
    \end{equation}
    where the first equality follows from~\eqref{R_mnk}, the first inequality and the second equality follow from \eqref{SINR_bound_2} and the fact that $f(\cdot)$ is assumed to be nondecreasing, the third equality follows from \eqref{gamma1}, \eqref{gamma2}, and the fact that $f(\cdot)$ is assumed to be nondecreasing with $f(0)=0$, and the fourth equality follows from the fact that $r_{m,n}^{(k)}$ is not a function of $\tilde k$.

    The physical meaning of the bound in~\eqref{Rate_bound} is similar to the SINR bound in~\eqref{SINR_bound}--the rate bound is obtained by assuming that RB $n$ is used in all sectors, except at most one sector. When RB $n$ is blanked in more than one interfering sector, only the most dominant \revN{blanked} interferer is considered.
%
%
    }
    }

    By substituting~\eqref{Rate_bound} in~\eqref{ICICNPHard}, the optimization problem \revN{in~\eqref{ICICNPHard} can be tightly approximated for dominant interference environment} as
    \begin{subequations}
    \label{ICICINLP}
    \begin{align}
    \underset{\substack{\revN{\{x^{(k)}_{m,n},I^{(k)}_{n}:} \\ \revN{k \in \mathcal{K}, m \in \mathcal{M}^{(k)}\}}}}
    {\text{maximize}}& \quad \sum_{k=1}^{K}\sum_{m=1}^{M^{(k)}} w^{(k)}_m x^{(k)}_{m,n} \Bigl( r^{(k)}_{m,n} +  \mathop {\max }\limits_{\tilde k \in \tilde{\cal K}^{(k)}} I_n^{(\tilde k)}{\tilde r}_{m,n}^{(k, \tilde k)} \Bigr)
    \label{ICICINLP:objective}\\
        \text{subject to}& \quad \sum_{m=1}^{M^{(k)}}x^{(k)}_{m,n}= 1-I^{(k)}_n, \quad \forall k, \label{ICICINLP:Constraint3}\\
        & \quad x^{(k)}_{m,n}, I^{(k)}_n \in \{0,1\}, \qquad \forall k, m.\label{ICICINLP:Constraint4}
    \end{align}
    \end{subequations}
The optimization problem \eqref{ICICINLP} is a non-linear binary integer optimization problem, which is in general, difficult to solve. As an intermediate step to reduce the complexity of solving \eqref{ICICINLP}, we convert it
to an equivalent binary linear optimization problem in the following section.}

\subsection{Transforming \eqref{ICICINLP} Into an Equivalent Binary Linear Optimization Problem}
\label{sec:BNLP2BLP}
Our approach to tackle the binary non-linear optimization problem given by~\eqref{ICICINLP} is to first transform it into an equivalent binary linear optimization problem. While binary linear optimization problems are still not easy to solve in general, good approximate solutions can be computed efficiently by relaxing the constraints that \revN{restrict the variables to be either 0 or 1, into weaker constraints that restrict the variables to assume any real value in the interval $[0,1]$}. The challenge is to construct an equivalent binary linear optimization problem that has a tight relaxation which means that the solution obtained by solving the relaxed version is very close to the one obtained by solving the binary linear optimization problem. This challenge is addressed in this section.


In order to convert the binary non-linear optimization problem given by~\eqref{ICICINLP} into an equivalent binary linear optimization problem, we need to convert the term $x^{(k)}_{m,n}\mathop {\max }\limits_{\tilde k \in \mathcal{K}^{(k)}} I_n^{(\tilde k)}{\tilde r}_{m,n}^{(k, \tilde k)}$ into a linear term. There are two sources of non-linearity in this term: the point-wise maximum and the multiplication.
The non-linear term can be written as
    \begin{equation}
    \label{BLP2LP:objective}
    x^{(k)}_{m,n} \max _{\tilde k \in \tilde{\cal K}^{(k)}} I_n^{(\tilde k)}{\tilde r}_{m,n}^{(k, \tilde k)}= \max_{y^{(k,\tilde k)}_{m,n} \in {\cal C}}
\sum_{\tilde k \in \tilde{\cal K}^{(k)}} y^{(k,\tilde k)}_{m,n} {\tilde r}_{m,n}^{(k, \tilde k)},
    \end{equation}
    where the variables $\{y^{(k,\tilde k)}_{m,n}: k \in \mathcal{K}, \tilde k \in \mathcal{K}^{(k)}, m \in \mathcal{M}^{(k)}, n \in \cal{N}\}$ are introduced as auxiliary variables that facilitate the conversion of the non-linear term into a linear term and


        \begin{equation}
    \label{FeasibleSet}
    \begin{array}{l l l}
    {\cal C}=\Big \{y^{(k,\tilde k)}_{m,n}:  
    &\displaystyle\sum_{\tilde k \in {\cal K}^{(k)}} y^{(k,\tilde k)}_{m,n} \leq 1, & \forall k, m,\\
    &y^{(k,\tilde k)}_{m,n} \leq x^{(k)}_{m,n}, &\forall k, \tilde k, m,\\
    & y^{(k,\tilde k)}_{m,n} \leq I^{(\tilde k)}_{n},  & \forall k, \tilde k, m, \\
    &y^{(k,\tilde k)}_{m,n} \in \{0,1\}, &\forall k, \tilde k, m \Big \}.
    \end{array}
    \end{equation}

    To see the equivalence, we note that since all variables assume binary values, then the point-wise maximum in the originally non-linear term is captured \revN{\label{objective_answer}in~\eqref{BLP2LP:objective}} and the first inequality in~\eqref{FeasibleSet}. Moreover, the multiplication in the originally non-linear term is captured in the second and the third inequalities in~\eqref{FeasibleSet}.

    By replacing the term $x^{(k)}_{m,n}\mathop {\max }\limits_{\tilde k \in \tilde{\cal K}^{(k)}} I_n^{(\tilde k)}{\tilde r}_{m,n}^{(k, \tilde k)}$ in~\eqref{ICICINLP:objective} with \eqref{BLP2LP:objective} and~\eqref{FeasibleSet}, we get an equivalent binary linear optimization problem. Unfortunately, we found experimentally that such an equivalent optimization problem leads to loose linear relaxation, i.e., it results in solutions that are far from the binary optimal solutions.
    As a result, we seek tighter equivalent formulations.

    A general approach to get tighter relaxation is by adding additional constraints that are called valid constraints~\cite[p. 585]{Bertsekas_book}. These valid constraints do not change the set of feasible binary solutions; however, with proper choice of these valid constraints, one can obtain tighter relaxations. Based on this approach, we construct a set ${\cal C}^\prime$ that is equivalent to ${\cal C}$ for the considered binary optimization problem, by adding two valid constraints:
    \begin{equation}
    \label{validConstraints}
    \begin{array}{l}
   \sum_{\tilde k \in \tilde{\cal K}^{(k)}} y^{(k,\tilde k)}_{m,n} \leq x^{(k)}_{m,n}, \quad \forall k, m,\\
    \sum_{m=1}^{M^{(\tilde k)}} y^{(k,\tilde k)}_{m,n} \leq I^{(\tilde k)}_{n}, \qquad \forall k, \tilde k.
    \end{array}
    \end{equation}
    Hence, ${\cal C}^\prime$ is given by
    \begin{equation}
    \label{FeasibleSet2}
    \begin{array}{l l l}
    {\cal C}^\prime=&\Big \{y^{(k,\tilde k)}_{m,n}:  
    \sum_{\tilde k \in \tilde{\cal K}^{(k)}} y^{(k,\tilde k)}_{m,n} \leq x^{(k)}_{m,n}, & \forall k, m,\\
    &\sum_{m=1}^{M^{(\tilde k)}} y^{(k,\tilde k)}_{m,n} \leq I^{(\tilde k)}_{n}, &\forall k, \tilde k,\\
    &y^{(k,\tilde k)}_{m,n} \in \{0,1\}, &\forall k, \tilde k, m\Big\}.
    \end{array}
    \end{equation}
Note that the constraints $y^{(k,\tilde k)}_{m,n} \leq x^{(k)}_{m,n}$ and
    $y^{(k,\tilde k)}_{m,n} \leq I^{(\tilde k)}_{n}$, given originally in \eqref{FeasibleSet},
    are omitted from the definition of  ${\cal C}^\prime$ given in \eqref{FeasibleSet2} since these constraints define a subset of the first two constraints given in \eqref{FeasibleSet2}.
    The equivalence of ${\cal C}$ and ${\cal C}^\prime$ is provided in the following Lemma.
    \begin{lemma}
    The sets ${\cal C}$ and ${\cal C}^\prime$ given by \eqref{FeasibleSet} and \eqref{FeasibleSet2}, respectively, are equivalent.
    \end{lemma}
    \begin{IEEEproof}
    \revNN{See Appendix~\ref{appendix_lemma1}.}
    \end{IEEEproof}

By replacing the term  $x^{(k)}_{m,n}\mathop {\max }\limits_{\tilde k} I_n^{(\tilde k)}{\tilde r}_{m,n}^{(k, \tilde k)}$ in \eqref{ICICINLP} with \eqref{BLP2LP:objective} and \eqref{FeasibleSet2}, we get the following binary linear optimization problem\footnote{\revNN{\label{removing_max}Note that the $\max$ operator in the right-hand side of~\eqref{BLP2LP:objective} is omitted from the objective of the optimization problem in~\eqref{ICICILP}, since the optimization problem is already a maximization problem.}}:

        \begin{subequations}
    \label{ICICILP}
    \begin{align}
    \underset{\substack{\{x^{(k)}_{m,n},y^{(k,\tilde k)}_{m,n},I^{(k)}_{n}: \\ k \in {\cal K}, \tilde k \in {\cal K}^{(k)}, m \in {\cal M}^{(k)\}}}}
    {\text{maximize}}& \quad \sum_{k=1}^{K}\sum_{m=1}^{M^{(k)}} w^{(k)}_m  \Bigg( x^{(k)}_{m,n} r^{(k)}_{m,n}  \nonumber\\
    &\qquad + \sum_{\tilde k \in {\cal K}^{(k)}} y^{(k,\tilde k)}_{m,n} {\tilde r}_{m,n}^{(k, \tilde k)} \Bigg)   \label{ICICILP:objective}\\
        \text{subject to} \qquad & \quad \sum_{m=1}^{M^{(k)}}x^{(k)}_{m,n}= 1-I^{(k)}_n,  \quad  \forall k,  \label{ICICILP:Constraint1}\\
        & \quad \sum_{\tilde k \in {\cal K}^{(k)}}y^{(k,\tilde k)}_{m,n} \leq x^{(k)}_{m,n},  \quad \forall k, m, \label{ICICILP:Constraint2}\\
        & \quad \sum_{m=1}^{M^{(k)}}y^{(k,\tilde k)}_{m,n} \leq I^{(\tilde k)}_{n}, \quad  \forall k, \tilde k, \label{ICICILP:Constraint3}\\
        & x^{(k)}_{m,n},y^{(k,\tilde k)}_{m,n}, I^{(k)}_n\in \{0,1\}, \quad \forall k, \tilde k, m. \label{ICICILP:Constraint4}
    \end{align}
    \end{subequations}


We finally remark that many other equivalent binary optimization problems can be formulated, e.g., by replacing the term $x^{(k)}_{m,n}\mathop {\max }\limits_{\tilde k} I_n^{(\tilde k)}{\tilde r}_{m,n}^{(k, \tilde k)}$ in \eqref{ICICINLP} with \eqref{BLP2LP:objective} and \eqref{FeasibleSet}. However, different equivalents will have different relaxations. Unlike many other equivalent formulations, \revN{the relaxed version of the formulation in~\eqref{ICICILP} has the following advantages:
    \begin{itemize}
        \item The optimal solution is \revN{provably} close to binary as we will show in Section~\ref{sec:Relaxation}.
        \item It can be readily solved in distributed manner using primal decomposition as we will show in Section~\ref{sec:decomposition}.
        \item It can be solved efficiently using network flow optimization tools as we will show in Section~\ref{sec:MCNF}.
         \end{itemize}
}

\subsection{Linear Optimization Relaxation}
\label{sec:Relaxation}
An upper bound on the optimum value of~\eqref{ICICILP} can be obtained by solving the relaxed version of~\eqref{ICICILP} which can be constructed by
replacing~\eqref{ICICILP:Constraint4} with the following constraints
    \begin{equation}
    \label{LP_variables_bounds}
    x^{(k)}_{m,n},y^{(k,\tilde k)}_{m,n}, I^{(k)}_n\in [0,1], \quad \forall k \in {\cal K }, \tilde k \in \tilde{\cal K}^{(k)}, m \in {\cal M }^{(k)}.
    \end{equation}
In particular, let $p^{\star}_{\textrm{Binary}}$ denote the optimal value of~\eqref{ICICILP}, let $p^{\star}_{\textrm{Relaxed}}$ denote the optimal value of the relaxed version of~\eqref{ICICILP}, and let $\hat p^{\star}_{\textrm{Relaxed}}$ denote the value of the objective function evaluated at a rounded-solution of the relaxed problem, such that the rounded solution is a feasible binary solution of~\eqref{ICICILP}. Then, we have the following inequalities
	\begin{equation}
	\label{optimalvalues}
\begin{array}{l}
	p^{\star}_{\textrm{Relaxed}} \geq p^{\star}_{\textrm{Binary}} \geq \hat p^{\star}_{\textrm{Relaxed}},
\end{array}
	\end{equation}
where the first inequality follows from the fact that the feasible set of the relaxed version is always a superset of the original problem and the second equality follows directly from the optimality of $p^{\star}_{\textrm{Binary}}$. We define the optimality gap, $\Delta_{\textrm{Opt}}$, in percentage as
	\begin{equation}
	\label{Eqn:OptimalityGap}
	\begin{array}{l l}
	\Delta_{\textrm{Opt}}&=   \bigl( {p^{\star}_{\textrm{Binary}} - \hat p^{\star}_{\textrm{Relaxed}}} \bigr) / {p^\star_{\textrm{Binary}}} \revN{\cdot100\%} \\
    &\leq \bigl( {p^{\star}_{\textrm{Relaxed}} - \hat p^{\star}_{\textrm{Relaxed}}} \bigr)/ {p^\star_{\textrm{Relaxed}}} \revN{\cdot100\%} ,
	\end{array}
	\end{equation}
where the inequality follows from~\eqref{optimalvalues}. \revN{Thus, one can compute an estimate on the optimality gap in polynomial time by solving a linear optimization problem. As we will show in Section~\ref{optimality_gap_results} through extensive simulations, by solving the relaxed problem and rounding the solution to the closest binary feasible solution, one can obtain a solution to~\eqref{ICICILP} that is near-optimal, i.e., with small $\Delta_{\textrm{Opt}}$.}
%

\revN{
An important objective measure of the tightness of a relaxation is the percentage of optimal relaxed variables that assume binary values. For instance, the optimality gap goes to zero as this percentage goes to 100\%. We now provide a closed-form theoretical guarantee on the this percentage. This theoretical guarantee is summarized in the following proposition. 
\begin{proposition}
\label{lemma:LP_relaxation}
The percentage of optimal variables of the relaxed version of~\eqref{ICICILP} that assume binary values is greater than or equal to $\frac{\tilde{K}(\bar{M}-1)}{(\tilde{K} + 1)\bar{M}+1} \cdot 100 \%$, where $\bar{M}$ is the average number of UTs per sector and $\tilde{K}$ is the number of neighboring interferers.
\end{proposition}
\begin{IEEEproof}
\rev{See Appendix~\ref{proof_lemma}.}
\end{IEEEproof}
For example, if $M^{(k)}=20, \forall k$, and $\tilde{K} = 6$, then using Proposition~\ref{lemma:LP_relaxation} we can deduce that the percentage of optimal variables of the relaxed version of~\eqref{ICICILP} that assume binary values is guaranteed to be greater than or equal to 80.8\%. 
}

However, solving the relaxed problem would require a central controller to be connected to all the BSs in order to solve a large linear optimization problem. Such a central controller is not supported in most contemporary cellular network standards, such as LTE, LTE-Advanced and IEEE~802.16m. Consequently, we seek \revN{in the following section} a distributed optimization method to solve the relaxed version of problem \eqref{ICICILP}. \ignore{Since most current and future standards (LTE, LTE-A, and IEEE 802.16m) do not support a central controller,
we seek a distributed optimization method to solve the relaxed version of problem \eqref{ICICILP}.}

\subsection{Primal Decomposition}
\label{sec:decomposition}
The relaxed version of~\eqref{ICICILP} has a special separable structure. In particular, for any set of fixed $\{I_n^{(k)}, \forall k \in {\cal K }\}$, it can be separated into $K$ optimization problems, each can be solved separately in each sector. \rev{In this section, we show how this structure is exploited to develop a distributed algorithm based on the primal-decomposition method\footnote{\label{prima_not_dual_answer}\revN{Primal-decomposition is considered in this paper instead of dual-decomposition since the structure is readily separable in the primal domain. Moreover, since the objective function is not strictly concave, recovering the optimal primal variables from optimal dual variables can be challenging for dual-decomposition method~\cite[p. 4]{PrimalDecomposition}}.}\cite[pp. 3--5]{PrimalDecomposition}.}

\rev{To exploit the separable structure}, let $\phi^{(k)}(I^{(1)}_n,\dots,I^{(K)}_n)$ denote the optimal value of the following optimization problem for given $\{I^{(1)}_n,\dots,I^{(K)}_n\}$:
    \begin{subequations}
    \label{ICICSubProblem}
    \begin{align}
    \underset{\substack{\revN{\{x^{(k)}_{m,n},y^{(k,\tilde k)}_{m,n}:} \\ \revN{m \in \mathcal{M}^{(k)},  \tilde k \in \tilde{\cal K}^{(k)}\}}}}
    {\text{maximize}}& \quad \sum_{m=1}^{M^{(k)}} w^{(k)}_m  \Bigl( x^{(k)}_{m,n} r^{(k)}_{m,n} + \sum_{\tilde k \in \tilde{\cal K}^{(k)}} y^{(k,\tilde k)}_{m,n} {\tilde r}_{m,n}^{(k, \tilde k)} \Bigr) \label{ICICSubProblem:objective}\\
        \text{subject to}& \qquad \sum_{m=1}^{M^{(k)}}x^{(k)}_{m,n}= 1-I^{(k)}_n, \label{ICICSubProblem:Constraint1}\\
        & \quad \sum_{\tilde k \in \tilde{\cal K}^{(k)}}y^{(k,\tilde k)}_{m,n} \leq x^{(k)}_{m,n}, \qquad \forall m, \label{ICICSubProblem:Constraint2}\\
        & \quad \sum_{m=1}^{M^{(k)}}y^{(k,\tilde k)}_{m,n} \leq I^{(\tilde k)}_{n}, \quad  \qquad \forall \tilde k, \label{ICICSubProblem:Constraint3}\\
        & \quad x^{(k)}_{m,n},y^{(k,\tilde k)}_{m,n}, \in [0,1], \qquad \forall m, \tilde k. \label{ICICSubProblem:Constraint4}
    \end{align}
    \end{subequations}
For reasons that will become apparent, we call~\eqref{ICICSubProblem} subproblem $k$.
Using~\eqref{ICICSubProblem}, the relaxed version of~\eqref{ICICILP} is equivalent to
    \begin{subequations}
    \label{ICICMaster}
    \begin{align}
    \underset{\substack{I^{(k)}_n,  k \in {\cal K }}}
    {\text{maximize}}& \quad \sum_{k=1}^{K} \phi^{(k)}(I^{(1)}_n,\dots,I^{(K)}_n)\label{ICICMaster:objective}\\
        \text{subject to}& \quad I^{(k)}_n\in [0,1], \quad \forall k \in {\cal K }. \label{ICICMaster:Constraint}
    \end{align}
    \end{subequations}
We call~\eqref{ICICMaster} the master problem. Therefore, the relaxed version of~\eqref{ICICILP} has been decomposed into a master problem, given by~\eqref{ICICMaster}, and $K$ subproblems,  each is given by~\eqref{ICICSubProblem} and can be solved separately in each sector.

{\label{subgradient_answer1}
\revN{
 The master problem given in~\eqref{ICICMaster} is a convex optimization problem with a concave objective function of the variables $\{I^{(1)}_n,\dots,I^{(K)}_n\}$ and a convex constraint set, since the relaxed version of~\eqref{ICICILP} is a linear (thus convex) optimization problem (cf.~\cite[p. 2]{PrimalDecomposition}). Since the objective is not necessarily differentiable, the projected-subgradient method can be used to solve this problem iteratively~\cite[p. 16]{SubgradientMethod}.
 }}
In each iteration, $K$ subproblems are solved in each sector in order to evaluate $\phi^{(k)}(I^{(1)}_n,\dots,I^{(K)}_n), \forall k \in {\cal K },$ and subgradients $[\Lambda^{\star (1)}_n,\dots,\Lambda^{\star (K)}_n]  \in \partial \sum_{k=1}^{K} \phi^{(k)}(I^{(1)}_n,\dots,I^{(K)}_n)$, where $\partial f(x)$ is the subdifferential of $f(\cdot)$ evaluated at $x$.
{\label{subgradient_answer2} \revN{
To explain how the subgradients $[\Lambda^{\star (1)}_n,\dots,\Lambda^{\star (K)}_n]$ are obtained, let $g^{(k)}_n (f(x))$ denote the $k^{\textrm{th}}$ element of a subgradient of the function $f(\cdot)$ evaluated at $x$ for RB $n$. Using the definition of a subgradient (cf.~\cite[p. 1]{Subgradients}), it is not difficult to see that $g^{(k)}_n$ is closed under addition, and thus,
    \begin{equation}
    \label{SubgradientFormula_pre}
    \begin{array}{l l}
    \Lambda^{\star (k)}_n & = g^{(k)}_n (\sum_{k=1}^{K} \phi^{(k)}(I^{(1)}_n,\dots,I^{(K)}_n)) \\
    & = \sum_{k=1}^{K} g^{(k)}_n(\phi^{(k)}(I^{(1)}_n,\dots,I^{(K)}_n)).
    \end{array}
    \end{equation}
    As shown in~\cite[p. 5]{PrimalDecomposition}, $g^{(k)}_n(\phi^{(k)}(I^{(1)}_n,\dots,I^{(K)}_n))$ can be obtained from an optimum lagrange multiplier (dual variable) that corresponds to the constraint where $I^{(K)}_n$ appears in its right-hand-side. Consequently, $\Lambda^{\star (k)}_n$ can be explicitly written as
     \begin{equation}
    \label{SubgradientFormula}
    \Lambda^{\star (k)}_n:=-\lambda^{\star (k)}_n + \sum_{\tilde k \in {\cal{K}}^{(k)}} \lambda^{\star (\tilde k,k)}_n ,
    \end{equation}
    where $\lambda^{\star (k)}_n$ is an optimum Lagrange multiplier  corresponding to constraint~\eqref{ICICSubProblem:Constraint1} and $\lambda^{\star (k,\tilde k)}_n, \tilde k \in {\cal{K}}^{(k)},$ are optimum Lagrange multipliers corresponding to constraints~\eqref{ICICSubProblem:Constraint3}.
    }


In order for each sector $k$ to calculate $\Lambda^{\star (k)}_n$, it requires the knowledge of $\lambda^{\star (k)}_n$, which can be obtained locally by solving~\eqref{ICICSubProblem}, and $\lambda^{\star (\tilde k,k)}_n, \tilde k \in {\cal{K}}^{(k)}$, which can be exchanged from the neighboring sectors. In other words, each sector $k$ sends $\lambda^{(k, \tilde k)}_n$ for all ${\tilde k}$ sectors that are in the neighborhood of sector $k$, for all $n$.
The master algorithm then updates its variables as
    \begin{equation}
    \label{I_k_n_initial}
    I^{(k)}_n:=I^{(k)}_n+\delta \Lambda^{(k)}_n, \forall k \in {\cal K },
    \end{equation}
 where $\delta$ is the step-size which can be chosen using any of the standard methods given in~\cite[pp. 3--4]{SubgradientMethod}. Then, each $I^{(k)}_n$ is projected into the feasible set of $[0,1]$ as follows
    \begin{equation}
    \label{Projection}
    I^{(k)}_n:=\left\{ \begin{array}{c c c}
    0,& \quad           &I^{(k)}_n \leq 0,\\
    I^{(k)}_n,& \quad 0<&I^{(k)}_n<1,\\
    1,& \quad            &I^{(k)}_n \ge 1.
    \end{array}\right.
    \end{equation}

    \revN{\label{RB_restirction_dist} Using~ \eqref{SubgradientFormula}, ~\eqref{I_k_n_initial},~and~\eqref{Projection},   each sector $k$ can compute the variables $\{I^{(k)}_n, \forall n\in\cal{N}\}$ in a distributed manner without the need for a centralized entity.} \revN{Then, each} sector $k$ exchanges $I^{(k)}_n$ with its neighbors and the process is repeated for $N_{\textrm{iter}}$ iterations. \revN{After that,} each $I^{(k)}_n$ is rounded to the nearest binary value which is denoted by $I^{\star (k)}_n$. Once $\{I^{\star (k)}_n\}$ are determined, local scheduling decision \revN{variables $\{x_{m,n}^{\star (k)}\}$ can be determined} in each sector $k$ separately as follows.
For every $m \in \mathcal{M}^{(k)}, n \in \mathcal{N}$, $R_{m,n}^{(k)}$ is calculated using~\eqref{SINR}, \eqref{R_mnk}, and $I^{\star (k)}_n$. To ensure feasibility of the resulting solution to problem~\eqref{ICICILP}, the scheduling decision variables are calculated as
            \begin{equation}
            \label{localx}
            \begin{array}{l}
             x_{m,n}^{\star (k)} = \left\{ {\begin{array}{*{20}{c}}
           {1,} & {m = \arg \mathop {\max }\limits_m {w^{(k)}_m R_{m,n}^{(k)}} \textrm{ and } I^{\star (k)}_n=0,} \\
           {0,} & {m \ne \arg \mathop {\max }\limits_m {w^{(k)}_m R_{m,n}^{(k)}} \textrm{ or } I^{\star (k)}_n=1}. \\
            \end{array}} \right. \\
             \end{array}\
             \end{equation}

{\label{subgradient_answer3}
\revN{Since the master problem is a convex optimization problem, t}}he subgradient algorithm is guaranteed to converge to the optimum solution of the relaxed version of problem~\eqref{ICICILP} as $N_{\textrm{iter}}\rightarrow\infty$ if $\delta$ is chosen properly~\cite[p. 6]{SubgradientMethod}. In this paper, we choose $\delta$ to be square summable but not summable by setting $\delta=c/p$, where $c > 0$ is a constant and $p$ is the iteration index.  This choice of \revN{$\delta$} guarantees convergence to the optimal solution as $N_{\textrm{iter}} \rightarrow \infty$~\cite[p. 6]{SubgradientMethod}. However, in \revN{practice} the algorithm must terminate after a finite number of iterations which raises the following  question: How
  different is the value obtained using finite iterations as compared to the true optimum obtained by solving~\eqref{ICICILP}? We address this question in Section~\ref{sec:SimulationResults} and show that few iterations are sufficient to achieve near-optimality.

  {
  \revN{
  The choice of the step-size, $\delta$, also affects the rate of message exchange (overhead) indirectly, at it affects $N_{\textrm{iter}}$ that is needed to achieve a particular optimality gap. {\label{relationshipdeltaNiter}\revNN{Since finding an explicit relationship between $N_{\textrm{iter}}$ and the step-size $\delta$ for subgradient algorithms is still an open research problem, tuning the step-size $\delta$ is performed empirically by choosing the constant $c$ to reduce $N_{\textrm{iter}}$ and thus reduce the rate of message exchange.}} Through extensive simulations, we found that a good choice of the parameter $c$ is mainly dependant on the choice of the desired utility. Thus, for a given utility, the parameter $c$ can be tuned offline only once before executing the algorithm. The relationship between $N_{\textrm{iter}}$  and the rate of message exchange required for the proposed algorithm will be discussed in Section~\ref{sec:complexity}.
  \label{step_size_overhead}
  }
  }

Clearly, the proposed algorithm relies heavily on solving the subproblem given by~\eqref{ICICSubProblem}. Hence, it is imperative to solve~\eqref{ICICSubProblem} as efficiently as possible. Interestingly, the subproblem given by~\eqref{ICICSubProblem} has a special network flow structure which can be exploited to devise efficient algorithms to solve it, as explained in the following section.
\rev{
\subsection{Transforming~\eqref{ICICSubProblem} into an Equivalent MCNF Problem}
\label{sec:MCNF}

The optimization problem~\eqref{ICICSubProblem} is a linear optimization problem which can be solved using generic simplex or interior-point methods. Nevertheless, we show in this section that~\eqref{ICICSubProblem} has a special network structure which makes it amenable to powerful network flow optimization methods that surpass conventional simplex and interior-point methods. In particular, we show that~\eqref{ICICSubProblem} can be converted into an equivalent MCNF optimization problem.

  An MCNF optimization problem is defined as finding a least cost way of sending certain amount of flow over a network that is specified by a directed graph of $v$ vertices and $e$ edges. Such a problem has $e$ variables, which represent the amount of flow on each arc, and $v$ linear equality constraints, which represent the mass-balance in each vertex, such that every variable appears in exactly two constraints: one with a coefficient of ${+}1$ and one with a coefficient of ${-}1$~\cite[p. 5]{Network_flows}. In addition to the mass-balance constraints, constraints on the lower and upper bounds on the amount of flow on each arch are also specified. The objective function is a weighted sum of the flows in each arc, where the weight is the cost per unit flow on that arc.
Thanks to the network structure of these problems, efficient combinatorial algorithms exist to solve such problems in strongly polynomial time, much faster than generic linear optimization solvers~\cite[Ch. 10]{Network_flows}.} For example, the enhanced capacity scaling algorithm can solve an MCNF problem with $v$ vertices and $e$ edges in $O\big( e\log v ( e+ v \log  v) \big)$\cite[p. 395]{Network_flows}.

\revN{\label{confusiong_MCNF} Although the problem in its original form given in~\eqref{ICICSubProblem} is not an MCNF optimization problem,  it can be transformed into an equivalent MCNF optimization problem as follows.} If we multiply both sides of constraint~\eqref{ICICSubProblem:Constraint3} with ${-}1$, we obtain a linear optimization problem with the following properties. Each variable appears in at most one constraint with a coefficient of ${+}1$ and at most one constraint with a coefficient of ${-}1$. According to Theorem~9.9 in~\cite[p. 315]{Network_flows}, a linear optimization problem with such a structure can be transformed into an equivalent MCNF optimization problem. To perform such transformation, we introduce the slack variables $\{s_m \geq 0, \forall m \in \mathcal{M}^{(k)}\}$ and surplus variables $\{\tilde s^{(\tilde k)} \geq 0, \forall \tilde k \in \tilde{\cal K}^{(k)}\}$ to convert the inequality constraints~\eqref{ICICSubProblem:Constraint2} and~\eqref{ICICSubProblem:Constraint3}, respectively, into equality constraints.
To obtain the mass-balance constraints, we also introduce a redundant constraint by summing constraints~\eqref{ICICMCNF:Constraint1}--\eqref{ICICMCNF:Constraint3}. In addition, we convert the maximization into minimization by negating the objective function.
Incorporating these \revN{transformations} into~\eqref{ICICSubProblem}, we get the following MCNF optimization problem:
        \begin{subequations}
    \label{ICICMCNF}
    \begin{align}
        \underset{\substack{\{x^{(k)}_{m,n},y^{(k,\tilde k)}_{m,n}, s_m,\tilde s^{(\tilde k)}:\\ m \in \mathcal{M}^{(k)},  \tilde k \in \tilde{\cal K}^{(k)}\}}}
    {\text{minimize}} \quad &-\sum_{m=1}^{M^{(k)}} w^{(k)}_m  \Bigl( x^{(k)}_{m,n} r^{(k)}_{m,n} \nonumber \\
    &\qquad + \sum_{\tilde k \in {\cal K}^{(k)}} y^{(k,\tilde k)}_{m,n} {\tilde r}_{m,n}^{(k, \tilde k)} \Bigr) \label{ICICMCNF:objective}\\
        \text{subject to}\quad &\sum_{m=1}^{M^{(k)}}x^{(k)}_{m,n}= 1-I^{(k)}_n, \label{ICICMCNF:Constraint1}\\
        \quad &\sum_{\tilde k \in {\cal K}^{(k)}}y^{(k,\tilde k)}_{m,n} -x^{(k)}_{m,n} +s_m = 0, \quad \forall m, \label{ICICMCNF:Constraint2}\\
        \quad -&\sum_{m=1}^{M^{(k)}}y^{(k,\tilde k)}_{m,n} - \tilde s^{(\tilde k)} = -I^{(\tilde k)}_{n}, \quad  \forall \tilde k,  \label{ICICMCNF:Constraint3}\\
        \quad -&\sum_{m=1}^{M^{(k)}}s_m + \sum_{\tilde k \in {\cal K}^{(k)}}\tilde s^{(\tilde k)} = -1+I^{(k)}_n \nonumber\\
        \quad & \phantom{\sum_{m=1}^{M^{(k)}}s_m + \sum_{\tilde k \in {\cal K}^{(k)}}\tilde s^{(\tilde k)}} \quad +\sum_{\tilde k \in {\cal K}^{(k)}} I^{(\tilde k)}_{n},\label{ICICMCNF:Constraint4}\\
        \quad & x^{(k)}_{m,n},y^{(k,\tilde k)}_{m,n},s_m,\tilde s^{(\tilde k)} \in [0,1], \quad \forall m , \tilde k. \label{ICICMCNF:Constraint5}
    \end{align}
    \end{subequations}
To see that~\eqref{ICICMCNF} is indeed an MCNF optimization problem, we note that the objective function is linear and the equality constraints \eqref{ICICMCNF:Constraint1}-\eqref{ICICMCNF:Constraint4} represent mass-balance constraints because each variable appears in two constraints: one with a coefficient of ${+}1$ and one with a coefficient of ${-}1$~\cite[p. 5]{Network_flows}.


\subsection{Pseudocode}
\label{sec:Pseudo}
A pseudocode of the proposed algorithm to be executed in every sector $k$ is given below.
    \begin{algorithm}
        \caption{Proposed ICIC algorithm to be executed in every sector $k$}\label{ICICPseudocode}
    \begin{algorithmic}[1]
    \REQUIRE $H_{m,n}^{(k,\tilde k)}$ and $w^{(k)}_m$, $\forall {\tilde k}\in\tilde{\cal K}^{(k)}, m \in \mathcal{M}^{(k)}, n \in \mathcal{N}$
    \ENSURE $x^{\star (k)}_{m,n}$, $I^{\star (k)}_n,$ $\forall m \in \mathcal{M}^{(k)}, n \in \mathcal{N}$
    \STATE Initialize  $I^{(k)}_n, \forall n \in \mathcal{N}$
    \STATE \textbf{Preprocessing:} Obtain $r_{m,n}^{(k)}$ and $\tilde r_{m,n}^{(k,\tilde k)}$ using \eqref{gamma1} and \eqref{gamma2}, $\forall {\tilde k}\in\tilde{\cal K}^{(k)}, m \in \mathcal{M}^{(k)}, n \in \mathcal{N}$.
       \FOR{$p = 1$ to $N_{\textrm{iter}}$}
            \STATE  \textbf{Solve a subproblem:} Obtain $x^{\star (k)}_{m,n}$,$y^{\star (k,\tilde k)}_{m,n}$,$\lambda^{\star (k)}_n$, and $\lambda^{\star (k, \tilde k)}_n$, $\forall {\tilde k} \in \tilde{\cal K}^{(k)}, n \in \mathcal{N},$ by solving the MCNF optimization problem \eqref{ICICMCNF}, $\forall n \in \mathcal{N}$.
            \STATE \textbf{Exchange subgradients:}
            Send $\lambda^{\star (k, \tilde k)}_n$  to sectors ${\tilde k}\in\tilde{\cal K}^{(k)}$, $\forall n \in \mathcal{N}$.
            \STATE \textbf{Update $I^{(k)}_n,$ $\forall n \in \mathcal{N}$:}
                \INDSTATE $\Lambda^{\star (k)}_n:=-\lambda^{\star (k)}_n + \sum_{\tilde k \in\tilde{\cal K}^{(k)}} \lambda^{\star (\tilde k,k)}_n $.
                \INDSTATE \textbf{Subgradient step:}
        				$I^{(k)}_n:=I^{(k)}_n+\delta \Lambda^{\star (k)}_n$
                \INDSTATE \textbf{Projection:} Project $I^{(k)}_n$ into the feasible set \eqref{Projection}.

            \STATE \textbf{Exchange $I^{(k)}_n$:} Send $I^{(k)}_n$ to sectors ${\tilde k}\in\tilde{\cal K}^{(k)}$, $\forall n \in \mathcal{N}$.
    \ENDFOR
    \STATE \textbf{Round the solution:} $I^{\star (k)}_n:=\lfloor I^{(k)}_n+0.5 \rfloor$, $\forall n \in \mathcal{N}$.
    \STATE \textbf{Local scheduling decisions:}
        \INDSTATE Obtain \ignore{$\Gamma_{m,n}^{(k)}$ and }$R_{m,n}^{(k)}$ by substituting $H_{m,n}^{(k,\tilde k)}$ and $I^{\star (k)}_n$ in \eqref{SINR} and \eqref{R_mnk}, $\forall m \in \mathcal{M}^{(k)}, n \in \mathcal{N}$.
        \INDSTATE Obtain $x^{\star (k)}_{m,n}$ by substituting $w^{(k)}_m$ and $R_{m,n}^{(k)}$ in \eqref{localx}, $\forall m \in \mathcal{M}^{(k)}, n \in \mathcal{N}$.
    \end{algorithmic}
    \end{algorithm}

    The proposed algorithm, whose pseudocode is given in Algorithm~\ref{ICICPseudocode}, has the following intuitive interpretation. Each sector $k$ estimates the benefit of shutting its own power on RB $n$ to neighboring sectors by $\sum_{\tilde k \in\tilde{\cal K}^{(k)}} \lambda^{\star (\tilde k,k)}_n $ and the benefit of using RB $n$ in sector $k$ by $\lambda^{\star (k)}_n$; if the former is more (less) than the latter,
    then sector $k$ increases (decreases) its soft decision on $I^{(k)}_n$ proportionally to the difference of these benefits, and possibly set it to 1 (0).



\subsection{Complexity Analysis of the Proposed Algorithm}
\label{sec:complexity}
In the following, we show that the worst-case time complexity of the proposed algorithm implemented in each sector $k$ is polynomial. We start by analyzing the complexity of solving the MCNF problem given by \eqref{ICICMCNF}. A graph that is equivalent to \eqref{ICICMCNF} is represented by \revN{$2 + \tilde{K} + M^{(k)}$} vertices, which is equal to the number of mass-balance constrains, and \revN{$(\tilde{K}+2)M^{(k)}+\tilde{K}$} arcs, which is equal to the number of variables. Thus, \revN{for fixed $\tilde{K}$}, \eqref{ICICMCNF} can be solved using the enhanced capacity scaling algorithm in $O\bigl( \bigl(M^{(k)}\log M^{(k)} \bigr)^2 \bigr)$\cite[p. 395]{Network_flows}. Since \eqref{ICICMCNF} is solved for each RB in $N_{\textrm{iter}}$ iterations, then the complexity of the proposed algorithm is $O\bigl( N_{\textrm{iter}}N \bigl(M^{(k)}\log M^{(k)} \bigr)^2 \bigr)$. Since $N_{\textrm{iter}}$ is a fixed constant, the complexity of the algorithm in sector $k$ as a function of $M^{(k)}$ and $N$ is $O\bigl( N \bigl(M^{(k)}\log M^{(k)} \bigr)^2 \bigr)$.



To reduce the information exchange between sectors, low complexity variants of the proposed algorithm can be implemented. One such implementation is to execute the algorithm every $\rho$ sub-frames. Each sector has time to implement the algorithm and calculate the $\{I^{(k)}_n\}$ values during $\rho$ sub-frames and send it to other BSs. The design parameter $\rho$ can be adjusted to suit the practical limitations. \ignore{The channel information could be filtered to represent the average channel information for the next $\rho$ sub-frames.
Also, a pipelined implementation is possible, where the algorithm is executed every sub-frame but with a delayed decision of $\tau$ sub-frames.}

\revN{
\label{message_exchange_distributed}
We now quantify the rate of message exchange required between sector $k$ and the neighboring $\tilde{K}$ sectors, in order to execute the proposed algorithm. For each iteration required by the proposed algorithm, each sector sends to each neighboring sector $N$ subgradients values and $N$ $I$ values. Thus, assuming $L_q$ bits is used to quantize the subgradients and $I$ values, 
then it is not difficult to see that the rate of message exchange in bit/sec is given by
\begin{equation}
\label{rate_exchange_proposed}
R_{\textrm{distributed exchange}}=\frac{2 N_{\textrm{iter}} \tilde{K} N L_q}{\rho \times 1 \textrm{ms}} \textrm{ bit/sec},
\end{equation}
 where the proposed algorithm is executed every $\rho$ sub-frames, each spans 1 ms.

We remark that the number of iterations required for the algorithm to converge is a constant parameter, $N_{\textrm{iter}}$, that can be configured to obtain the desired tradeoff between complexity and the optimality gap, where increasing $N_{\textrm{iter}}$ can potentially reduces the optimality gap at the expense of higher computational complexity and signalling overhead, as we will demonstrate in Section~\ref{optimality_gap_results}.


}

\revN{
\subsection{Advantages of Distributed Implementation}
\label{advantages_distributed}
We conclude this section by highlighting three main advantages of the proposed distributed scheme as compared to centralized implementation of the same scheme:
\begin{enumerate}[I.]
\item \ignore{An important advantage of the proposed scheme is that} It does not require a centralized entity to collect the channel information from all UTs in the network and make network-wide scheduling decisions. This particular advantage makes the proposed algorithm applicable to current and emerging standards such as LTE and LTE-advanced, where a centralized entity is not supported.

\item The proposed algorithm can potentially reduce the message exchange  as compared to a centralized implementation. To elaborate, we start by quantifying the rate of message exchange required for centralized implementation between each sector $k$ and a centralized entity and then compare it with the rate of message exchange for the distributed scheme given in~\eqref{rate_exchange_proposed}. Since each sector $k$ is required to send to the centralized entity $N \times M^{(k)} (\tilde{K} + 1)$ channel coefficients and assuming $L_q$ bits is used to quantize the channel coefficients, 
then it is not difficult to see that the rate of message exchange for a centralized implementation in bit/sec is given by
\begin{equation}
\label{rate_exchange_centralized}
R_{\textrm{centralized exchange}}=\frac{N \times M^{(k)} (\tilde{K} + 1) L_q}{\rho \times 1 \textrm{ms}} \textrm{ bit/sec}.
\end{equation}
Using \eqref{rate_exchange_proposed} and \eqref{rate_exchange_centralized}, the ratio of the rate of message exchange required by the centralized scheme to the rate of message exchange by the proposed scheme is given by
\begin{equation}
\label{ratio_rate_exchange_centralized}
\frac{R_{\textrm{centralized exchange}}}{R_{\textrm{distributed exchange}}}=\frac{M^{(k)} (\tilde{K} + 1)}{2 (\tilde{K}) N_{\textrm{iter}}}.
\end{equation}
Thus, as long as $N_{\textrm{iter}} < \frac{1}{2} M^{(k)} (1 + \frac{1}{\tilde{K}}) $, then the proposed distributed implementation requires lower rate of message exchange than the centralized implementation. For instance, if $M^{(k)} = 20$ UTs, $\tilde{K} = 6$ interferers, and $N_{\textrm{iter}} = 5$, then, the proposed scheme requires 2.33 times lower rate of message exchange than the centralized implementation.

\item The amount of computations that is needed to be performed in a centralized entity will scale at least linearly with the number of sectors, $K$. Thus, scaling the network would require to scale also the computational capabilities of the centralized entity. However, this is not an issue for the proposed distributed implementation since the computations required are distributed among the $K$ sectors and the computations needed by each sector is independent of the number of sectors in the network.
\end{enumerate}
}

\ignore{
\subsection{Enhanced ICIC in Heterogenous Networks}
\label{sec:eICIC}
Although the proposed algorithm is presented in the context of homogenous network deployment, where the access network consists of macro BSs, it can also work in the context of heterogenous network as an enhanced ICIC (eICIC) scheme~\cite{eICIC}. In a heterogenous network (also called multi-tier network), user terminals may be granted access through a macro-BS, pico-BS, femto-BS, or through a fixed relay station. Pico-BSs, femto-BSs, and relays are deployed inside the coverage region of the macro-BS and they may reuse the resources used by the macro-BS. As a result, a very harsh interference environment is created which raises the need for powerful ICIC schemes. The proposed algorithm can be still used in heterogenous network as long as communication is possible between the different access points types, which can be realized in practice. For example, in 3GPP Release 10 (LTE-Advanced)~\cite{LTERel10}, macro-pico BS communication is possible through the X2 interface and macro-femto BS communication is possible through the OAM (Operations, Administration, and Maintenance) interface~\cite{eICIC}. The proposed algorithm can be executed every $\rho$ sub-frames, where $\rho$ can be adjusted to suit the practical limitations of X2 and OAM interfaces.
}

\section{Simulation Setup and Parameters}
\label{sec:SimulationSetup}
The simulation parameters are based on the IMT-Advanced Urban macro-cell (UMa) scenario~\cite{IMTAdvanced} which are provided in Table~\ref{table:SimulationParameters}.
Based on the IMT-Advanced guidelines, a hexagonal layout with wrapround is considered with $57$ hexagonal sectors and 10 UTs per sector. These sectors are  served by $19$ BSs, each with a tri-sector antenna to serve a 3-sector cell-site. 
Monte Carlo simulations are carried 
over  $1000$ sub-frames and averaged over 10 independent drops.
In each drop, 
the average received power from all sectors are calculated for each UT; this involves the calculation of the pathloss, correlated shadowing, and antenna gains. 
Variations of the received signal power due to small-scale  fading within  each RB is negligible and thus, the  channels are assumed  fixed over each RB. For other RBs, the channels  assume different  values depending on  the time-frequency correlation  of the IMT-Advanced model for the UMa scenario~\cite{IMTChannelModel}. We use the AMC strategy given in \revN{Table~\ref{table:AMCTable}}~\cite{RS_WICOM10}. Channel-aware scheduling and ICIC are done afterwards, on a sub-frame by sub-frame basis. After scheduling the last sub-frame, the time-average throughput for each user is calculated. Then, another drop commences and the process repeats. Finally, time-averaged users throughput from all drops are saved for further processing and plotting. Further details of the simulation procedure are given in~\cite[Section 7]{IMTAdvanced}. The simulator is \revN{validated} with the UMa simulation results provided in~\cite{WinnerCalibration}.

  \begin{table}[t]
    \centering
    \caption[The AMC table]{The AMC table which is used to determine the achievable data rate on each RB for each user based on the SINR. This table is based on the AMC strategy given in~\cite{RS_WICOM10}.}
    \revN{
  	\begin{tabular}{|c | c|| c | c|}
      \hline
  	RB SINR & Data rate & RB SINR & Data rate \\
  range (dB) & per RB (kbit/sec) & range (dB) & per RB (kbit/sec) \\
  \hline \hline
    ($-\infty$ -6.1] & 0 & (5.8, 8.5] & 291.6 \\     \hline
    (-6.1, -4.1] & 35.3 & (8.5, 9.9] & 388.4 \\     \hline
    (-4.1, -2.0] & 56.4 & (9.5, 12.5] & 418.3 \\     \hline
    (-2.0, -0.2] & 92.4 & (12.5, 14.8] & 544.3\\     \hline
    (-0.2, 1.9] & 131.4 & 14.8. 16.1] & 648.1\\     \hline
    (1.9, 3.8] & 177.4 & (16.1, 17.8] & 721.7\\     \hline
    (3.8, 5.8] & 223.1 & (17.8, $\infty$) & 807.4\\     \hline
    \end{tabular}
        }
    \label{table:AMCTable}
    \end{table}

	\begin{table}
    \caption{\rev{Simulation Parameters based on IMT-Advanced UMa scenario.}}
    \centering
    \scalebox{0.9}{
	\begin{tabular}{|c | c|}
    \hline
	  Parameter & Assumption \\ \hline \hline
	  Number of sectors & 57 (wraparound) \qquad\\ 	 \hline
	  Number of UTs per sector & 10 \\ \hline
          Inter-site distance & 500 m\\     \hline
          BS height & 25 m\\       \hline
          Min. dist. between a UT and a BS & 25 m\\ \hline
          UT speed & 30 km/h \\ \hline
	  Bandwidth (downlink)& 10 MHz\\ \hline 
          Sub-carrier spacing & 15 KHz\\ \hline
          Number of RBs ($N$)& 50 \\ \hline
          OFDM symbol duration & 66.67 \textmu s \\ \hline
	  Number of sub-carriers per RB& 12\\ \hline
          Number of OFDM symbols per RB& 7\\ \hline
                   Number of drops & 10 \\ \hline
         Number of sub-frames per drop  & 1000 \\ \hline
		Noise power per RB ($P_N$) & -114.45 dBm  \\ \hline
          Carrier frequency  & 2.0 GHz\\     \hline
           Total BS transmit power & 46 dBm\\   \hline
         Path loss and shadowing & Based on UMa scenario \cite{IMTAdvanced}\\         \hline
         Averaging window ($t_{c}$) & 100 \\ \hline
         BS antenna gain (boresight)& 17 dBi\\ \hline
         User antenna gain & 0 dBi\\ \hline
         Feeder loss & 2 dB \\ \hline
         Channel estimation delay & 4 sub-frames\\ \hline
         SINR estimation margin  & 6 dB \\
        \hline
          BS antenna tilt $(\phi_{t})$  & $12^\circ$ \cite{WinnerCalibration} \\ \hline
           BS horizontal  antenna pattern & $A(\theta)=\scalebox{0.75}[1.0]{\(- \)}\min\left[12(\frac{\theta}{70^\circ})^2, 20 \textrm{ dB}\right] $\ignore{\cite{IMTAdvanced}} \\ \hline
		   BS elevation antenna pattern & $A_{e}(\phi)=\scalebox{0.75}[1.0]{\(- \)}\min\left[12(\frac{\phi - \phi_{t}}{15^\circ})^2, 20 \textrm{ dB}\right]$ \ignore{\cite{IMTAdvanced}} \\ \hline
	       BS combined  antenna pattern & $ \scalebox{0.75}[1.0]{\(- \)}\min\left[\scalebox{0.75}[1.0]{\(- \)}\left( A(\theta)+A_{e}(\phi)
\right), 20 \textrm{ dB}\right]$ {\cite{IMTAdvanced}}\\ \hline
   Small-scale fading model & IMT-Advanced channel model \cite{IMTChannelModel}\\ \hline
	     Traffic model & Full buffer \\ \hline

	 \end{tabular}
}
\label{table:SimulationParameters}
	\end{table}

\subsection{Importance of Accurate Simulation}
\label{sec:importance_of_accurate_simulation}
Although explaining how to do accurate simulations is not the main purpose of this paper, we nevertheless highlight an aspect in the simulation that has important effect on assessing the performance of ICIC schemes. In particular, we demonstrate that the way UTs are associated to sectors has a significant impact on the simulation results. In the following, we explain three association strategies, namely, wideband SINR-based, wideband SINR-based (excluding shadowing), and geographical-based association. Wideband SINR is defined as the ratio of the average power received from the serving sector to the sum of the average power received from all other sectors and the noise power at the UT (i.e., small scale fading is not included in the calculation of the wideband SINR). In wideband SINR-based association, each UT is associated with the antenna sector to which it has the highest wideband SINR. This association strategy resembles reality, provides the most favorable results, and it is widely used by evaluation groups. A consequence of this strategy is that the coverage region of each sector is not hexagonal and it changes from drop to drop due to the different shadowing realizations. Another consequence is that a UT may not be associated to the closest sector antenna as it may experience heavy shadowing to that sector antenna. A more convenient way of doing simulation is to exclude shadowing in the calculation of the wideband SINR which leads to fixed coverage regions for each sector in all drops. Due to the directional antenna patterns, the coverage regions of each sector is not hexagonal. In geographical-based association, a UT is associated to a particular sector antenna if it resides inside the hexagonal area of that sector. In this strategy, the coverage region of each sector is hexagonal.

In Fig.~\ref{Fig:SINRCDF}, we plot the CDF of the wideband SINR using the three association strategies for reuse-1. For \revN{validation} purposes, we also include the average CDF results produced by seven WINNER+ partners using different simulation tools for UMa scenario~\cite{WinnerCalibration}. It is clear from the figure that wideband SINR-based association strategy produces a CDF that agree very well with the calibrated results; however, the other two association strategies produce a heavy tail which would impact the throughput of the users at the cell edge. The heavy tail is a direct consequence of associating UTs to sectors in a suboptimal manner. As a result, these two association strategies may not be suitable for assessing the performance of ICIC schemes as they tend to exaggerate the gains achieved for UTs at the cell-edge. This is the case because the cell-edge user throughput for reuse-1 for these two association strategies is very low. This makes any improvement in cell-edge user throughput to be large as compared to the very low values of reuse-1. Consequently, we use wideband SINR-based association strategy to assess the performance of ICIC schemes.

{
\label{validation}
\revN{We conclude this section by highlighting the main motivations for performing SINR validation in system simulations:
\begin{enumerate}
\item System simulators are complex in nature which make them error-prone. As a result, careful verification of the simulation results is imperative.
\item Validation of the simulation makes it easy for other researchers 
    to compare the performance of our schemes with other schemes in a widely-accepted scenario such as the UMa scenario specified by IMT-advanced.
\item As we observed in Fig.~\ref{Fig:SINRCDF}, the gains of ICIC schemes may be exaggerated if the simulation assumptions, such as the UT-BS association strategy, does not resemble reality. Validation of the SINR distribution is a good way to verify the validity of the simulation assumptions.
\end{enumerate}
}
}

    \begin{figure}[!ht]
    \centering
    \psfrag{Wideband SINR (dB)}[][][1.2]{dB}
    \psfrag{Probability (Wideband SINR < Abscissa)}[][][1.2]{Probability (Wideband SINR $<$ Abscissa)}
    \psfrag{Wideband SINR-based}[][][1.1]{\quad Wideband SINR-based}
    \psfrag{Wideband SINR-based2}[][][1.1]{\ \ Wideband SINR-based}
    \psfrag{(excluding shadowing)}[][][1.1]{\quad (excluding shadowing)}
    \psfrag{Geographically-based}[][][1.1]{\ \quad Geographically-based}
    \psfrag{RRReference results (WINNER+)}[][][1.1]{\ \quad Reference results (WINNER+)}
    \resizebox{.5\textwidth}{!}
    {\includegraphics[clip = true,trim = 0.7cm 0cm 0.7cm 0cm]{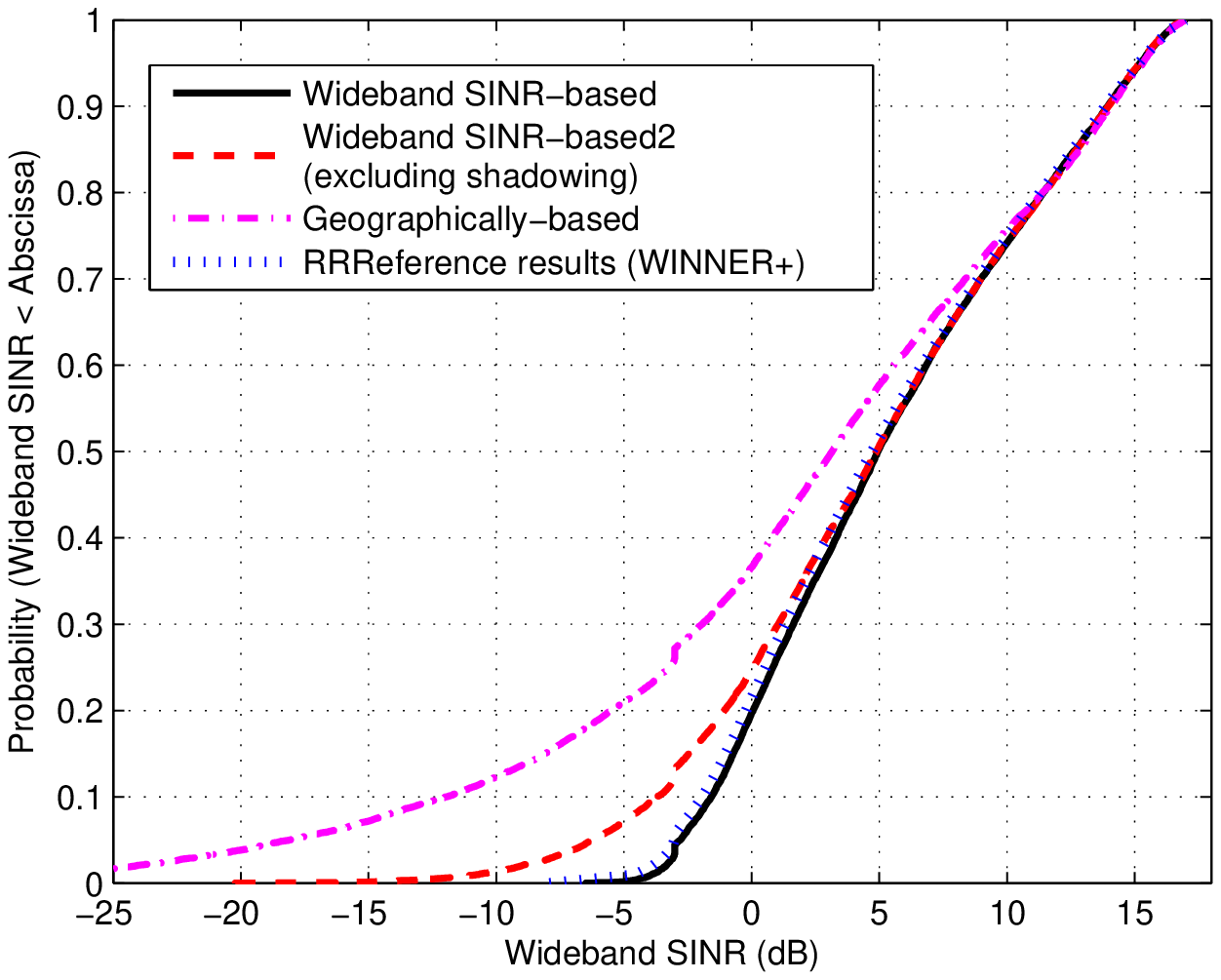}}

    \caption{CDF of the wideband SINR for different association strategies. Reference results~(WINNER+) refers to the average CDF results produced by seven WINNER+ partners \cite{WinnerCalibration}.}
    \label{Fig:SINRCDF}
    \end{figure}

\subsection{Baseline Schemes}
\label{BaselineScheme}
\ignore{Although the main purpose of the paper is to develop distributed and near-optimal ICIC scheme, we also compare the proposed scheme to a number of baseline schemes, to demonstrate the benefit of developing near-optimal as compared to developing ICIC algorithms based on heuristics.}
In this paper, the followings are used as baseline schemes:
reuse-1, reuse-3, partial frequency reuse~(PFR)~\cite{Sternad2003}, dynamic FFR~\cite{Hussain2009}, and optimum FFR~\cite{VTC2008_OptimumFFR}. While reuse-1, reuse-3, and PFR are static schemes, dynamic and optimum FFR are implemented as dynamic schemes.

In PFR, RBs are divided such as 30 RBs (inner band) are used in all sectors (reuse-1) while 20 RBs (outer band) are shared among sites in a classical reuse-3 pattern. More details about this algorithm can be found in \cite{Sternad2003}.

While dynamic FFR and optimum FFR were designed to maximize the sum-rate, they can be easily modified to accommodate maximizing weighted sum-rates, and they work with differentiable and non-differentiable AMC functions, which make them good baseline schemes to be compared with the proposed scheme. For fair comparison, we convert the constraint on instantaneous minimum rate (used in dynamic and optimum FFR) to a constraint on average minimum rate, which can be implemented by updating the weights in every sub-frame according to the procedure given in \cite{Stolyar_PF}. We finally remark that both dynamic FFR and optimum FFR are centralized schemes.

In dynamic FFR, a centralized controller is used to determine which RBs belong to inner band (in reuse-1 pattern) and which RBs belong to outer bands (in reuse-3 pattern). The decision is made to maximize the total utility, assuming each RB is used by all the users in the region where this RB can be used, and summed for all the sectors. This assumption is critical to the development of the algorithm. Then, scheduling is done locally by each BS. More details about this algorithm can be found in \cite{Hussain2009}.

In optimum FFR, the number of RBs used in inner and outer bands, and the reuse factor of the outer band are determined optimally using a centralized controller, without considering channel fading. Without channel fading, all RBs seen by a particular user have the same SINR. This makes the optimization problem tractable. However, if channel fading is considered, then the computational complexity for optimum FFR is exponential. More details about this algorithm can be found in \cite{VTC2008_OptimumFFR}.

It is common for FFR schemes (e.g., PFR and optimum FFR) to divide users into two classes: inner (cell-center) users and outer (cell-edge) users, based on SINR or distance from BS. Inner users are restricted to use inner band while outer users are restricted to use outer band. We found such restriction degrades the performance of FFR and as such, this restriction is removed to realize the full potential of FFR schemes.
  This restriction is already removed from dynamic FFR~\cite{Hussain2009} for the same reason.

\section{\revN{Simulation and Analytical Results}}
\label{sec:SimulationResults}
\subsection{Optimality Gaps}
\label{optimality_gap_results}
  To develop an efficient algorithm to solve the difficult optimization problem given by~\eqref{ICICINLP} in a distributed manner, two sources of sub-optimality were introduced, namely, relaxing the integer constraints and solving the master optimization problem in finite iterations.
\revN{To understand the effect of these sources of sub-optimality, we present in Fig.~\ref{Fig:OptGap_convergence} the mean optimality gap~\eqref{Eqn:OptimalityGap}, the 5th percentile, and 95th percentile of the distribution of the optimality gap obtained from simulating 22,000 instances of optimization problems for \revNN{\label{citation_PIMRC2011}different numbers of UTs and four IMT-Advanced scenarios, namely, UMa, Urban micro-cell (UMi), Rural macro-cell (RMa), and Suburban macro-cell (SMa) scenarios}~\cite{ABS_PIMRC2011}. In the first sub-frame, a random initial point is used. In the subsequent sub-frames, the optimal solution of the previous frame is used as an initial point. As it can be seen from this figure, the proposed algorithm converges fast to low optimality gaps; in particular, executing the algorithm in 5 iterations provided less than 1.5\% mean optimality gap. As a result, the number of iterations, $N_{\textrm{iter}}$,  is chosen to be $N_{\textrm{iter}}=5$ throughout the paper.}

\revN{Fig.~\ref{Fig:OptGap_convergence} illustrates} that our algorithm can solve the bound optimization problem~\eqref{ICICINLP}, \revN{which is suitable in dominant interference environment}, in a near-optimum manner, \revN{as expected from Proposition~\ref{lemma:LP_relaxation}}. A natural question to ask is how different is the optimal value obtained by solving the bound optimization as compared to the optimal value obtained by solving the original strongly NP-hard problem~\eqref{ICICNPHard} using exhaustive search? Due to the exponential computational complexity of exhaustive search, simulating a system of 57 sectors is not feasible. As a result, we only show the optimality gap for a system of 12 sectors in Table~\ref{table:OptimalityGap}, for proportional-fair scheduling, i.e., $\alpha=1$. As we can see, the proposed scheme can achieve, on average, about 96\% of the optimum value achieved using exhaustive search, if it is executed once (5 iterations). One can also reduce the optimality gap further by executing the algorithm more than once. That is, after finding the optimum $\{I^{(k)}_n\}$ in one run, the algorithm sets $H_{m,n}^{(\tilde k,k)}=0, \   \forall m, n, k, \tilde k | I^{(k)}_n=1$, and executes the algorithm again. In this case, one can achieve about 97.6\% of the optimum value achieved using exhaustive search, i.e., an incremental gain, at the expense of more computational complexity. This suggests that the bound optimization is indeed a good method to achieve near-optimality and shows that most of the gain is already captured by executing the algorithm only once.

      \begin{table}
      \centering
      \caption{Mean and standard deviation of the optimality gap (\%) compared to the optimal value of the original problem \eqref{ICICNPHard}}

      \begin{tabular}{|c|c|c|}
    \hline
     Number of runs & Mean (\%)	& Standard deviation (\%) \\ \hline
    1 &	3.8 & 1.7 \\ \hline
    2 & 2.4 & 0.8	\\ \hline
      \end{tabular}
      \label{table:OptimalityGap}
      \end{table}

    \begin{figure}[!ht]
    \centering
     \psfrag{Number of Iterations}[cl][cl][1]{Number of Iterations}
     \psfrag{Optimality Gap (\%)}[cl][cl][1]{Optimality Gap (\%)}
     \psfrag{5th percentile}[cl][cl][1]{$95^{\textrm{th}}$ percentile}
     \psfrag{95th percentile}[cl][cl][1]{$5^{\textrm{th}}$ percentile}
     \psfrag{Mean optimality }[cl][cl][1]{Mean}
    \resizebox{.5\textwidth}{!}
    {\includegraphics[clip = true,trim = 0.7cm 0cm 0.7cm 0cm]{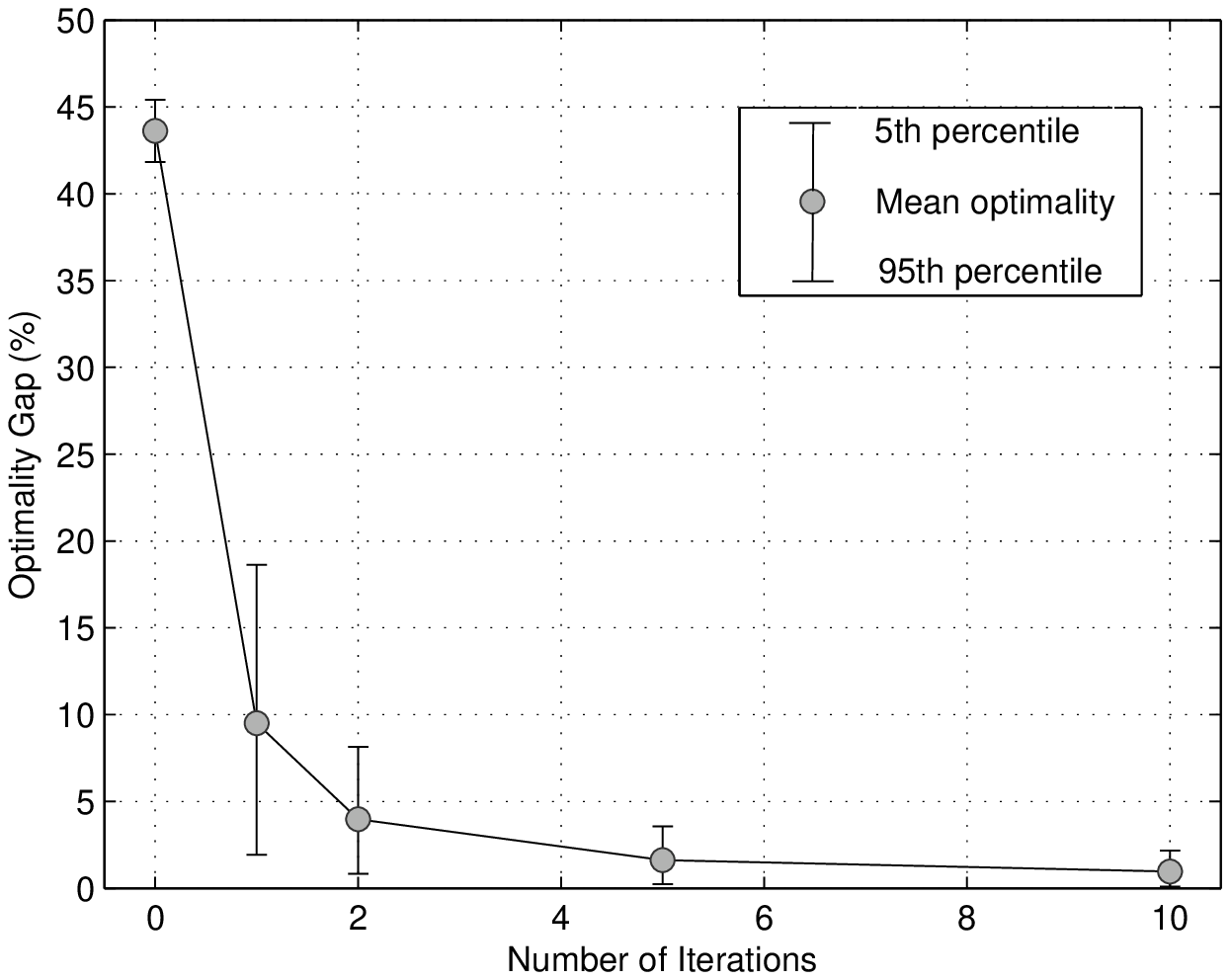}}
    \caption{Optimality gap (\%) as a function of the number of iterations.}
    \label{Fig:OptGap_convergence}
    \end{figure}



\subsection{Comparing the Performance of the Proposed Scheme with the Baseline Schemes}
In this section, we compare the performance of the proposed scheme with the baseline schemes presented in Section~\ref{BaselineScheme}. In Fig.~\ref{Fig:CDFMue1}, we show the CDF of the normalized time-average UT throughput for all schemes. Normalization is performed by dividing the user throughput over the total downlink bandwidth, which is $10$ MHz. In all schemes, $\alpha$-fair scheduler is used with fairness exponent of $\alpha=2$ (cf. Section~\ref{sec:ProblemStatement}). To facilitate the comparison, we define the normalized cell-edge and cell-center user throughputs as the $5^\textrm{th}$ and the $95^\textrm{th}$ percentiles of the normalized user throughputs, respectively. It is clear from the figure that reuse-1 has the worst cell-edge performance (0.0323 bit/sec/Hz) as compared to the other three schemes, due to the excessive interference experienced at the cell-edge. Reuse-3, PFR, dynamic FFR, optimum FFR, and the proposed scheme achieve normalized cell-edge user throughputs of 0.0391, 0.0417, 0.0412, 0.0435, and 0.0430 bit/sec/Hz, respectively. However, reuse-3, PFR, dynamic FFR, and optimum FFR improve the cell-edge performance at the expense of reducing the overall throughput, especially for UTs close to the cell-center. For example, the $95^\textrm{th}$ percentile achieved by reuse-3, PFR, dynamic FFR, and optimum FFR are 0.108, 0.136, 0.107, and 0.125 bit/sec/Hz, respectively, as compared to 0.156 and 0.154 bit/sec/Hz achieved by reuse-1 and the proposed scheme, respectively. Interestingly, the proposed scheme combines the advantages of all schemes, as it provides high cell-edge and  cell-center throughputs simultaneously. {\label{fifty_percentile_case} \revNN{In addition to the improvement in cell-center and cell-edge throughputs, the proposed scheme also outperforms the other schemes in terms of the normalized median throughput ($50^\textrm{th}$ percentile)}.} Indeed, the gain achieved by the proposed scheme increases for higher fairness exponent, $\alpha$, or higher $\bar R_{\min}$, as we will see shortly. 

    \begin{figure}[!ht]
    \centering
     \psfrag{Probability (Normalized user throughput< Abscissa)}[][][1]{Probability (Normalized user throughput $<$ Abscissa)}
     \psfrag{Normalized user throughput (bit/sec/Hz)}[][][1]{bit/sec/Hz}
     \psfrag{Reuse 1}[][][1]{Reuse-1}
     \psfrag{Reuse 3}[][][1]{Reuse-3}
     \psfrag{PFR}[][][1]{PFR}
     \psfrag{Dynamic FFR}[][][1]{ Dynamic FFR}
     \psfrag{Optimum FFR}[][][1]{ Optimum FFR}
     \psfrag{Proposed}[][][1]{Proposed}
    \resizebox{.5\textwidth}{!}
    {\includegraphics[clip = true,trim = 0.7cm 0cm 0.7cm 0cm]{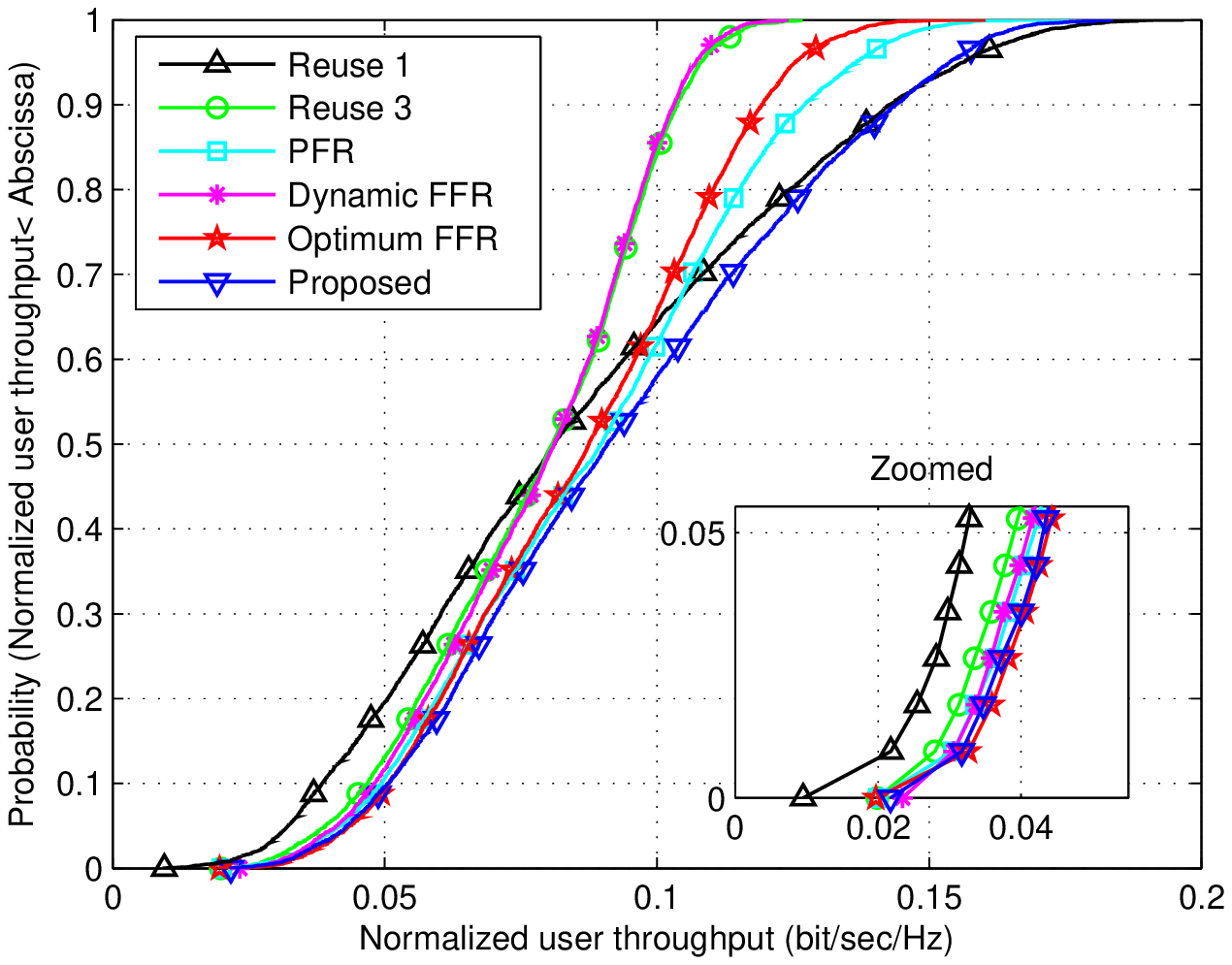}}
    \caption{CDF of the normalized user throughput of all UTs in the network for UMa scenario, $\alpha=2$ and $\bar R_{\min}=0$.}
    \label{Fig:CDFMue1}
    \end{figure}

\ignore{
mue=2
Cell-center Throughput for Reuse-1 is 0.1562 (bit/sec/Hz)
Cell-center Throughput for Reuse-3 is 0.1081 (bit/sec/Hz)
Cell-center Throughput for PFR is 0.1365 (bit/sec/Hz)
Cell-center Throughput for Dynamic FFR is 0.1071 (bit/sec/Hz)
Cell-center Throughput for Optimum FFR is 0.1254 (bit/sec/Hz)
Cell-center Throughput for Proposed is 0.1537 (bit/sec/Hz)
Cell-edge Throughput for Reuse-1 is 0.0323 (bit/sec/Hz)
Cell-edge Throughput for Reuse-3 is 0.0391 (bit/sec/Hz)
Cell-edge Throughput for PFR is 0.0417 (bit/sec/Hz)
Cell-edge Throughput for Dynamic FFR is 0.0412 (bit/sec/Hz)
Cell-edge Throughput for Optimum FFR is 0.0435 (bit/sec/Hz)
Cell-edge Throughput for Proposed is 0.0430 (bit/sec/Hz)

}

To further examine the performance of the different schemes, we show in Fig.~\ref{Fig:CellEdgeVsThroughputVsRmin} the normalized cell-edge user throughput and the normalized aggregate sector throughput for all schemes and for different minimum average rate requirements $\bar R_{\min}$, assuming proportional-fair scheduler, i.e., $\alpha=1$. As expected, the general trend for all schemes is that as $\bar R_{\min}$ increases, the sector throughput decreases and the cell-edge user throughput increases. For high $\bar R_{\min}$, we observe that reuse-3, PFR, optimum FFR, dynamic FFR, and the proposed scheme have significantly higher cell-edge throughput than reuse-1. On the other hand, for small $\bar R_{\min}$ it is clear that reuse-3, PFR, dynamic FFR, and optimum FFR incur significant loss in the aggregate sector throughput as compared to reuse-1. Interestingly, the proposed scheme performs very well in both the cell-edge and the sector throughput as compared to all other schemes for all values of $\bar R_{\min}$. We also plot similar tradeoff curves in Fig.~\ref{Fig:CellEdgeVsThroughputVsMue} by varying the fairness exponent $\alpha$ and fixing $\bar R_{\min}=0$, similar to~\cite{RS2011PIMRC}. Again, the proposed scheme outperforms other schemes in both the cell-edge and the aggregate sector throughput.


    \begin{figure}[!ht]
    \centering
          \psfrag{Normalized aggregate sector throughput (bit/sec/Hz)}[][][1.1]{Normalized aggregate sector throughput (bit/sec/Hz)}
     \psfrag{Normalized cell-edge user throughput (bit/sec/Hz)}[][][1.1]{Normalized cell-edge throughput (bit/sec/Hz)}
     \psfrag{Reuse 1}[][][1]{Reuse-1}
     \psfrag{Reuse 3}[][][1]{Reuse-3}
     \psfrag{PFR}[][][1]{PFR}
     \psfrag{Dynamic FFR}[][][1]{ Dynamic FFR}
     \psfrag{Optimum FFR}[][][1]{ Optimum FFR}
     \psfrag{Proposed}[][][1]{Proposed}
        \resizebox{.5\textwidth}{!}{\includegraphics[clip = true,trim = 0.3cm 0cm 0.4cm 0cm]{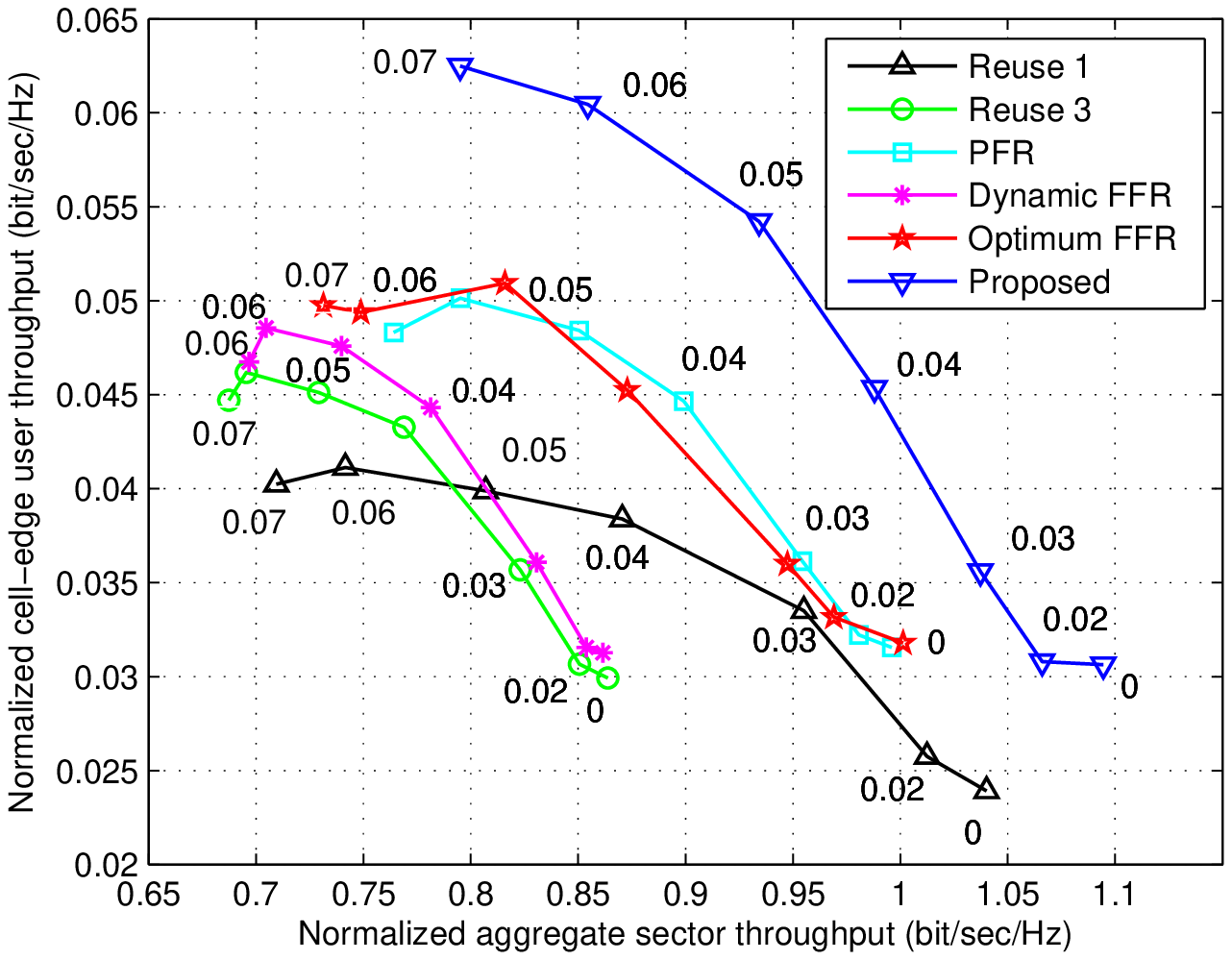}}
    \caption{Normalized cell-edge throughput versus normalized aggregate sector throughput for different schemes, for $\alpha=1$ and $\bar R_{\min} \in \{0,0.02,0.03,0.04,0.05,0.06,0.07\}$.}
    \label{Fig:CellEdgeVsThroughputVsRmin}
    \end{figure}

    \begin{figure}[!ht]
       \centering
     \psfrag{Normalized aggregate sector throughput (bit/sec/Hz)}[][][1.1]{Normalized aggregate sector throughput (bit/sec/Hz)}
     \psfrag{Normalized cell-edge user throughput (bit/sec/Hz)}[][][1.1]{Normalized cell-edge user throughput (bit/sec/Hz)}
     \psfrag{Reuse 1}[][][1]{Reuse-1}
     \psfrag{Reuse 3}[][][1]{Reuse-3}
     \psfrag{PFR}[][][1]{PFR}
     \psfrag{Dynamic FFR}[][][1]{ Dynamic FFR}
     \psfrag{Optimum FFR}[][][1]{ Optimum FFR}
     \psfrag{Proposed}[][][1]{Proposed}
    \resizebox{.5\textwidth}{!}{\includegraphics[clip = true,trim = 0.4cm 0cm 0.4cm 0cm]{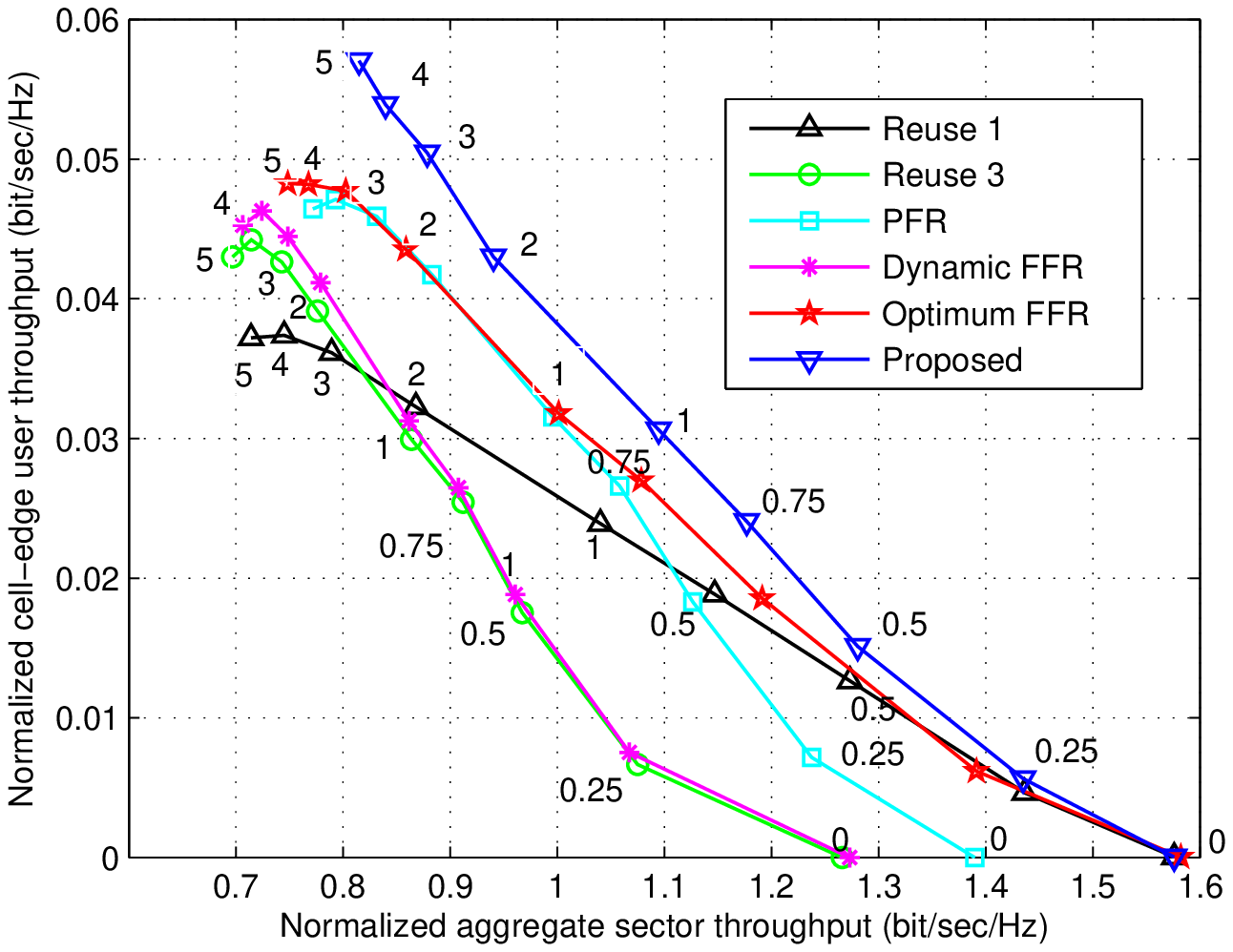}}
    \caption{Normalized cell-edge throughput versus normalized aggregate sector throughput for different schemes, for $\bar R_{\min}=0$ and $\alpha \in \{0,0.25,0.50,0.75,1,2,3,4,5\}$.}
    \label{Fig:CellEdgeVsThroughputVsMue}
    \end{figure}

\ignore{
To examine the performance of the different schemes, we show in Fig. \ref{Fig:CellEdgeVsThroughputVsMue} the normalized cell-edge user throughput and the normalized aggregate sector throughput for all schemes and for different fairness exponents, similar to \cite{RS2011PIMRC}. The general trend for all schemes (as expected) is that as $\alpha$ increases (high degree of fairness), the sector throughput decreases and the cell-edge user throughput increases. For high $\alpha$, we observe that reuse-3, PFR, and the proposed scheme have significantly higher cell-edge throughput than reuse-1. On the other hand, for small $\alpha$ it is clear that both reuse-3 and PFR incur significant loss in the aggregate sector throughput as compared to reuse-1. Interestingly, the proposed scheme performs very well in both the cell-edge and the sector throughput as compared to all other schemes for all values of $\alpha$. We remark that for the case of $\alpha=0$, which corresponds to max-SINR scheduler, the proposed scheme converges to reuse-1. This is the case because max-SINR scheduler favors cell-center UTs which usually receive minimal interference from neighboring sectors; thus, RB \revN{blanking} is not needed.}

In Fig.s~\ref{Fig:GainVsTh_VsRmin} and~\ref{Fig:GainVsCellEdge_VsRmin}, we take a closer look at the gains achieved by the different schemes as compared to reuse-1. In Fig. \ref{Fig:GainVsTh_VsRmin}, we plot the gains in cell-edge throughput achieved for a given aggregate sector throughput. The gain in cell-edge throughput for a particular scheme for a given sector throughput $x$ is given by $\textrm{Gain}_\textrm{scheme}(x)=\frac{\textrm{CellEdge}_\textrm{scheme}(x)-\textrm{CellEdge}_\textrm{reuse1}(x)}{\textrm{CellEdge}_\textrm{reuse1}(x)} \cdot 100\%$, where $\textrm{CellEdge}_\textrm{scheme}(x)$ is the cell-edge throughput achieved by a particular scheme at sector throughput $x$, which can be obtained from the tradeoff curves given in Fig.~\ref{Fig:CellEdgeVsThroughputVsRmin}. For a wide range of aggregate sector throughput, the proposed scheme achieves large gains ($50\%$ to $60\%$). Dynamic FFR, optimum FFR, and the proposed scheme lose some of the gain if it is executed every 10 sub-frames (as expected for any dynamic scheme); however, the gains for the proposed scheme are consistently better than the other schemes. Similarly, we plot the gains in aggregate sector throughput for a given cell-edge throughput in Fig.~\ref{Fig:GainVsCellEdge_VsRmin}. The proposed scheme achieves consistently higher gains in aggregate sector throughput than the other schemes, especially for high cell-edge throughput.

    \begin{figure}[!ht]
    \centering
    \psfrag{Normalized aggregate sector throughput (bit/sec/Hz)}[][][1.1]{Normalized aggregate sector throughput (bit/sec/Hz)}
     \psfrag{Gain \(\%\) in cell-edge user throughput}[][][1.1]{Gain (\%) in cell-edge user throughput}
     \psfrag{(rrrrrr=10 sub-frames)}[][][1]{($\rho=10$ sub-frames)}
     \psfrag{Reuse 1}[][][1]{Reuse-1}
     \psfrag{Reuse 3}[][][1]{Reuse-3}
     \psfrag{PFR}[][][1]{PFR}
     \psfrag{Dynamic FFR}[][][1]{ Dynamic FFR}
     \psfrag{Optimum FFR}[][][1]{ Optimum FFR}
     \psfrag{Proposed}[][][1]{Proposed}
                        \resizebox{.54\textwidth}{!}{\includegraphics[clip = true,trim = 1.5cm 0cm 0cm 0cm]{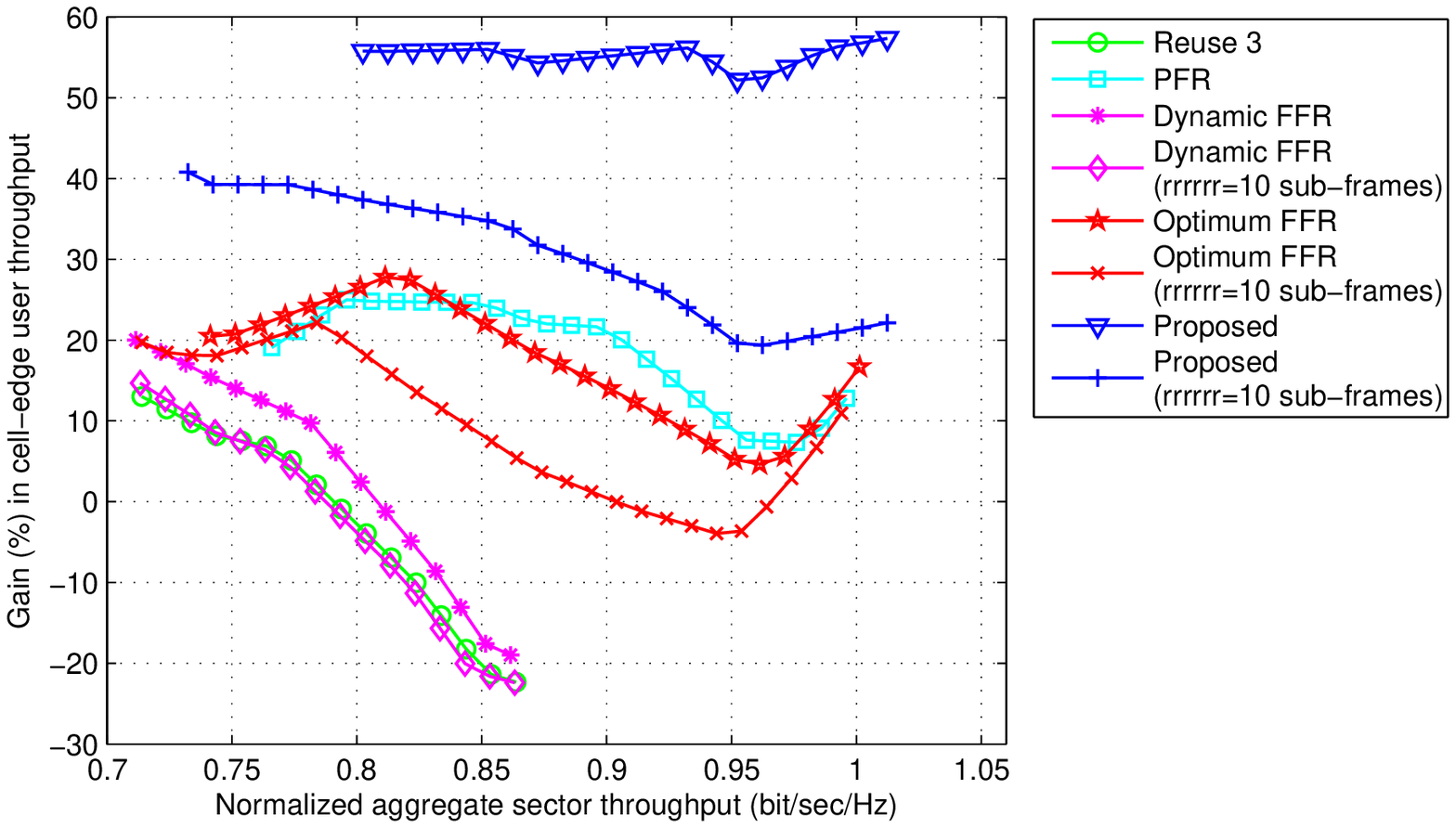}}
    \caption{Gain (\%) in cell-edge throughput versus aggregate sector throughput, as compared to reuse-1.}
    \label{Fig:GainVsTh_VsRmin}
    \end{figure}

    \begin{figure}[!ht]
    \centering
    \psfrag{Normalized cell-edge user throughput (bit/sec/Hz)}[][][1.1]{Normalized cell-edge user throughput (bit/sec/Hz)}
     \psfrag{Gain \(\%\) in aggregate sector throughput}[][][1.1]{Gain (\%) in aggregate sector throughput}
     \psfrag{Reuse 1}[][][1]{Reuse-1}
     \psfrag{Reuse 3}[][][1]{Reuse-3}
     \psfrag{PFR}[][][1]{PFR}
     \psfrag{Dynamic FFR}[][][1]{ Dynamic FFR}
     \psfrag{Dynamic FFR (rrrrrr=10 sub-frames)}[][][1]{  Dynamic FFR ($\rho=10$ sub-frames)}
     \psfrag{Optimum FFR}[][][1]{ Optimum FFR}
     \psfrag{Optimum FFR (rrrrrr=10 sub-frames)}[][][1]{  Optimum FFR ($\rho=10$ sub-frames)}
     \psfrag{Proposed}[][][1]{Proposed}
     \psfrag{Proposed (rrrrrr=10 sub-frames)}[][][1]{ Proposed ($\rho=10$ sub-frames)}
    \resizebox{.5\textwidth}{!}{\includegraphics[clip = true,trim = 0.3cm 0cm 0.7cm 0cm]{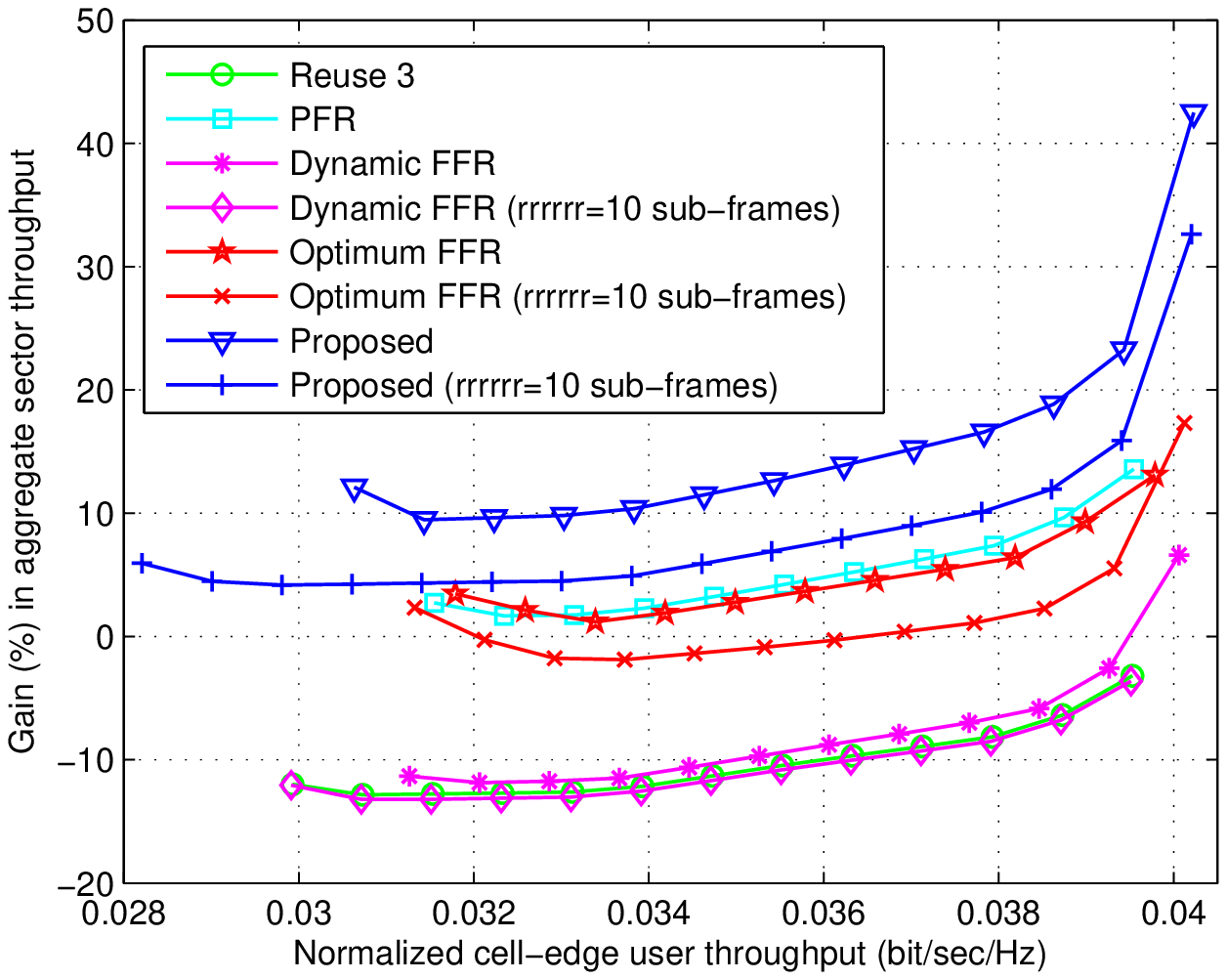}}
    \caption{Gain (\%) in sector throughput versus cell-edge throughput, as compared to reuse-1.}
    \label{Fig:GainVsCellEdge_VsRmin}
    \end{figure}


In Fig.~\ref{Fig:Outage}, we plot the outage probability, which is defined as the probability of having the average UT throughput less than $\bar R_{\min}$ for different schemes. It is clear from the figure that the proposed scheme achieves much lower outage probability as compared to other schemes. For example, at $\bar R_{\min}=0.05$ bits/sec/Hz, the proposed scheme has an outage probability that is at least 3.5 times less than those for PFR and optimum FFR, and at least 6 times less than those for reuse-1, reuse-3, and dynamic FFR.

    \begin{figure}[!ht]
    \centering
    \psfrag{Probability \(Normalized user throughput < Rmin \) \(\%\)}[b][][1]{Probability (Normalized user throughput $< \bar R_{\min}$)}
    \psfrag{Rmin \(bit/sec/Hz\)}[][][1]{$\bar R_{\min}$ (bit/sec/Hz)}
     \psfrag{Reuse 1}[][][1]{Reuse-1}
     \psfrag{Reuse 3}[][][1]{Reuse-3}
     \psfrag{PFR}[][][1]{PFR}
     \psfrag{Dynamic FFR}[][][1]{ Dynamic FFR}
     \psfrag{Optimum FFR}[][][1]{ Optimum FFR}
     \psfrag{Proposed}[][][1]{Proposed}

      \psfrag{5}[][][1]{0.05\quad}
      \psfrag{10}[][][1]{0.01  }
      \psfrag{15}[][][1]{0.15  }
      \psfrag{20}[][][1]{0.20  }
      \psfrag{25}[][][1]{0.25  }
      \psfrag{30}[][][1]{0.30  }
      \psfrag{35}[][][1]{0.35  }
      \psfrag{40}[][][1]{0.40 }

      \psfrag{0}[][][1]{0}
      \psfrag{0.01}[][][1]{0.01}
      \psfrag{0.02}[][][1]{0.02}
      \psfrag{0.03}[][][1]{0.03}
      \psfrag{0.04}[][][1]{0.04}
      \psfrag{0.05}[][][1]{0.05}
      \psfrag{0.06}[][][1]{0.06}

    \resizebox{.5\textwidth}{!}{\includegraphics[clip = true,trim = 0.4cm 0cm 0.4cm 0cm]{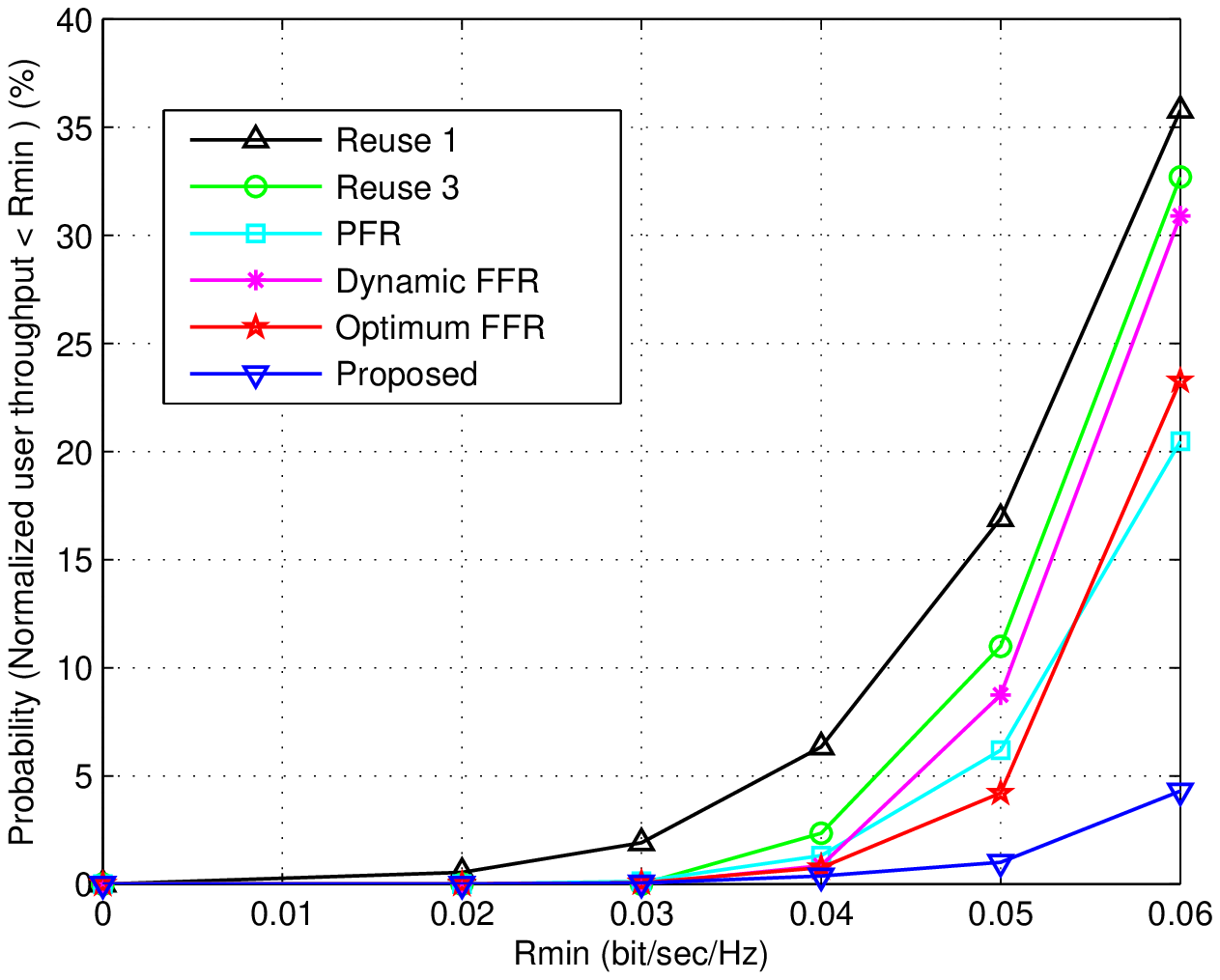}}
    \caption{Outage probability versus $\bar R_{\min}$, which is defined as the probability that the average UT throughput less than $\bar R_{\min}$.}
    \label{Fig:Outage}
    \end{figure}

\revN{
 \subsection{Comparing the Complexity of the Proposed Scheme with the Baseline Schemes}
 \label{sec:complexity_comparison}
 By analyzing the dynamic FFR algorithm in~\cite{Hussain2009} and the optimum FFR algorithm in~\cite{VTC2008_OptimumFFR}, we obtained the computational complexity and the rate of message exchange required to execute both algorithms. Due to space limitation, we only summarize the final result in Table~\ref{table:complexity}. In terms of computational complexity, it can be seen from Table~\ref{table:complexity} that the proposed algorithm scales better (worse) than dynamic and optimum FFR as $N$ or $K$ ($M^{(k)}$) increases. In terms of the rate of message exchange, the proposed scheme scales better than dynamic and optimum FFR as $M^{(k)}$ increases, while optimum FFR scales best as $N$ increases.
%
%

\begin{table}
\caption{\revN{Comparison between the proposed distributed scheme, dynamic FFR, and optimum FFR, in terms of computational complexity and rate of message exchange required}}
\label{table:complexity}
\revN{
\resizebox{.5\textwidth}{!}{
\begin{tabular}{|c|c|c|c|}
\hline
  Scheme & \multicolumn{2}{|c|}{Computational complexity} & Rate of message \\    
   &  \multicolumn{2}{|c|}{} & exchange required\\ \cline{2-3}
   & Centralized entity& Sector $k$  & for sector $k$ (bit/sec) \\ \hline
  Proposed Scheme                   & None    & $O\bigl(N \bigl(M^{(k)}\log M^{(k)} \bigr)^2 \bigr)$  & $\frac{2 N_{\textrm{iter}} \tilde{K} N L_q}{\rho \times 1 \textrm{ms}}$ \\ \hline
  Dynamic FFR~\cite{Hussain2009}          &  $O\bigl( N(N-1) K)$  & $O\bigl( N M^{(k)} )$ & $\frac{2N  M^{(k)} L_q}{\rho \times 1 \textrm{ms}}$ \\
    \hline
  Optimum FFR~\cite{VTC2008_OptimumFFR} & $O\bigl(N^2 \displaystyle\sum_{k=1}^K M^{(k)})$  &  $O\bigl( N M^{(k)} )$ & $\frac{ M^{(k)} L_q}{\rho \times 1 \textrm{ms}}$ \\
    \hline

\end{tabular}
}
}
\end{table}

%
%
%

}

\subsection{Statistics of the Average Number of \revN{Blanked} RBs}
In Fig.~\ref{Fig:NPDF}, we plot the probability mass function of the average number of \revN{blanked} RBs per sector for different fairness exponents. The proposed scheme has the flexibility to change the distribution of the \revN{blanked} resources according to the desired fairness level. As $\alpha$ increases, cell-edge users become more important and thus more resources need to be \revN{blanked}, and vice versa. The figure shows also that the proposed scheme acts as reuse-1 (no \revN{blanking}) and reuse-3 (2/3 of RBs are \revN{blanking}) with very small probabilities.

    \begin{figure}[!ht]
    \psfrag{Probability (Number of restricted RBs = Abscissa)}[][][1]{Probability (Number of \revN{blanked} RBs = Abscissa)}
    \psfrag{Number of restricted RBs}[][][1]{}
     \psfrag{alpha=1}[][][1.2]{$\alpha=1$}
     \psfrag{alpha=4}[][][1.2]{$\alpha=4$}
          \psfrag{alpha=10}[][][1.2]{$\alpha=10$}
    \psfrag{Reuse 1}[][][1]{Reuse-1}
     \psfrag{Reuse 3}[][][1]{Reuse-3}
     \psfrag{PFR}[][][1]{PFR}
    \centering
    \resizebox{.5\textwidth}{!}{\includegraphics[clip = true,trim = 0.4cm 0cm 0.4cm 0cm]{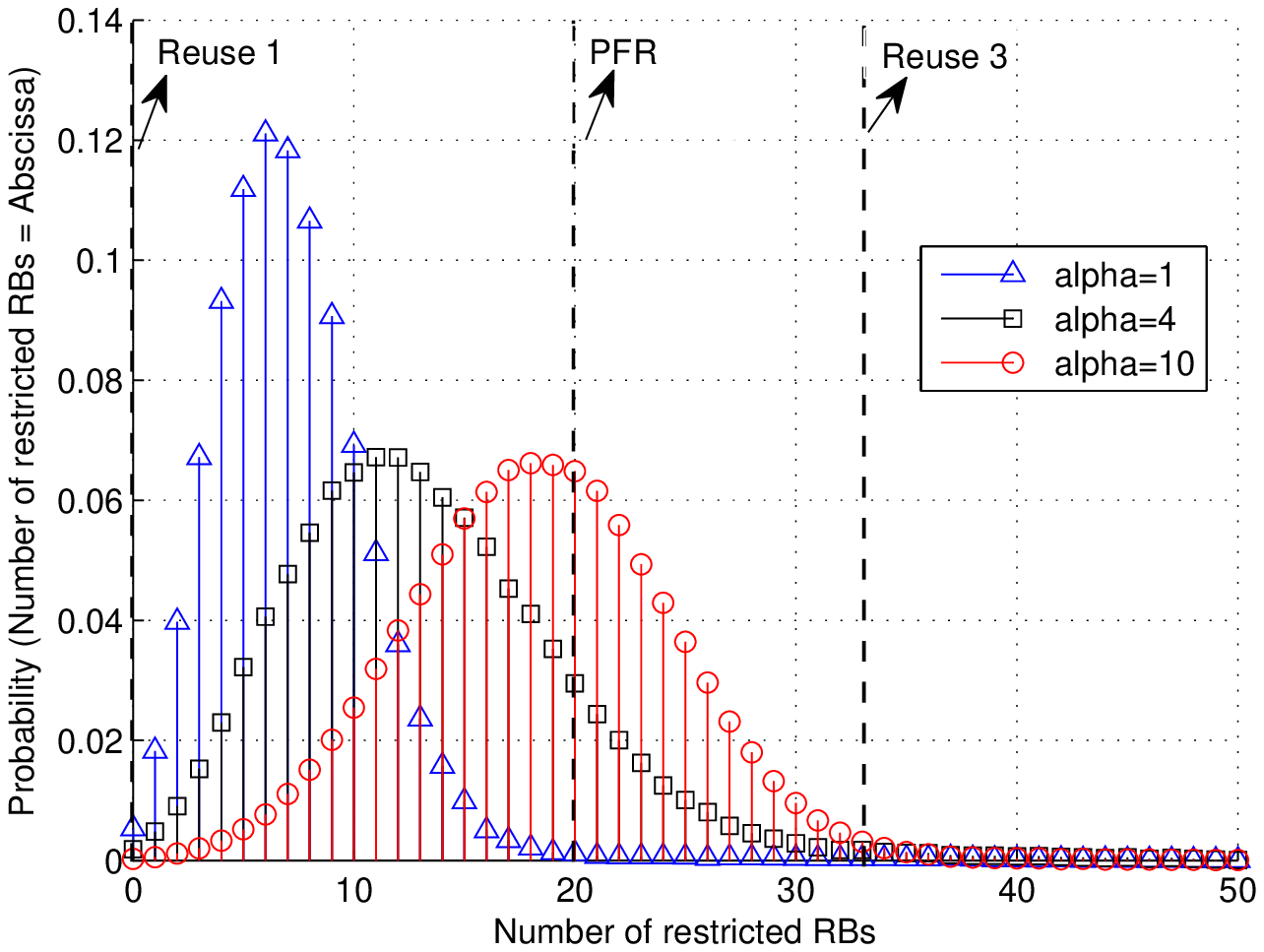}}
    \caption{Probability mass function of the average number of \revN{blanked} RBs per sector (the total number of RBs is 50). }
    \label{Fig:NPDF}
    \end{figure}

\section{Conclusions}
\label{sec:Conclusions}

\revN{In this paper, we tackled the problem of distributed  multi-cell resource allocations with blanking where the objective is to maximize the weighted sum-rate. This problem is known to be strongly NP-hard problem; nevertheless, we demonstrated that this problem can be tightly relaxed to a linear programming problem in a dominant interference environment, in which for each user the received power from each interferer is significantly greater than the aggregate received power from all other weaker interferers. Using linear relaxation, we proposed a polynomial-time distributed algorithm that is not only guaranteed to be tight for dominant interference environment, but which also computes an upper bound on the optimality gap.}
The proposed scheme is developed using the primal-decomposition method, which is utilized to decompose the problem into a master problem and multiple subproblems. The master problem is solved iteratively using the projected-subgradient method. We reveal that each subproblem has a network flow structure which makes it amenable to powerful MCNF algorithms and thus results in significant reduction in the computational complexity.
Simulation results show that 
the proposed scheme achieves high gain in aggregate throughout, cell-edge throughput, and outage probability, as compared to reuse-1, reuse-3, partial frequency reuse, dynamic fractional frequency reuse, and optimum fractional frequency reuse.
\revNN{\label{paper_decomposition_TSP2014} An interesting extension to this work is to use the decomposition algorithms proposed in~\cite{decomposition_TSP2014} to devise distributed ICIC algorithms.}

\section*{Acknowledgment}
\addcontentsline{toc}{section}{Acknowledgment} The authors would like to express their gratitude to Dr. Ho Ting Cheng  and Dr. Zhijun Chao from Huawei Technologies Canada, and Dr. Ramy Gohary from Carleton University, for their helpful comments and suggestions.

\appendices
\section{On the tightness of the bound given in \eqref{SINR_bound}}
\label{Appendix:Bound}
In this Appendix, we show that the SINR bound given by \eqref{SINR_bound} is tight in the sense if the bound is increased by $\Delta$, then the actual SINR will increase by at least $\Delta$. Let $\tilde k^\star=\arg\mathop {\max }\limits_{\tilde k \in \tilde{\cal K}^{(k)}}I_n^{(\tilde k)}H_{m,n}^{(k,\tilde k)}$, then the exact SINR expression given by \eqref{SINR} can be written as
        \begin{equation}
    \begin{array}{l}
    \Gamma_{m,n}^{(k)} = \frac{{{P_C}H_{m,n}^{(k,k)}}}{{{P_C}\left(\sum\limits_{\tilde k \neq k} {H_{m,n}^{(k,\tilde k)}}  -  I_n^{\tilde k^\star}H_{m,n}^{(k,\tilde k^\star)} -\sum\limits_{\tilde k \neq k,\tilde k \neq \tilde k^\star } {I_n^{(\tilde k)}H_{m,n}^{(k,\tilde k)}}\right)+ {P_N}}},\\
        \phantom{\Gamma_{m,n}^{(k)}} = \frac{{{P_C}H_{m,n}^{(k,k)}}}{{{P_C}\left(\sum\limits_{\tilde k \neq k } {H_{m,n}^{(k,\tilde k)}}  -  I_n^{\tilde k^\star}H_{m,n}^{(k,\tilde k^\star)} \right)+ {P_N}}}\times \\
        \phantom{\Gamma_{m,n}^{(k)}=}
        \left(1+\frac{P_C\sum\limits_{\tilde k \neq k,\tilde k \neq \tilde k^\star} {I_n^{(\tilde k)}H_{m,n}^{(k,\tilde k)}}}{{P_C}\sum\limits_{\tilde k \neq k} {\left(1-I_n^{(\tilde k)}\right)H_{m,n}^{(k,\tilde k)}}  + {P_N}}\right)\\
    \phantom{\Gamma_{m,n}^{(k)}} \geq \frac{{{P_C}H_{m,n}^{(k,k)}}}{{{P_C}\left(\sum\limits_{\tilde k \neq k} {H_{m,n}^{(k,\tilde k)}}  -  I_n^{\tilde k^\star}H_{m,n}^{(k,\tilde k^\star)} \right)+ {P_N}}},
    \end{array}
    \end{equation}
    where the right-hand side of the inequality is the bound given by \eqref{SINR_bound}. From the previous expression, if the bound is increased by $\Delta$, then $\Gamma_{m,n}^{(k)}$ will increase by $\Delta \times    \Bigl(1+\frac{P_C\sum\limits_{\tilde k = 1,\tilde k \neq k,\tilde k \neq \tilde k^\star}^K {I_n^{(\tilde k)}H_{m,n}^{(k,\tilde k)}}}{{P_C}\sum\limits_{\tilde k = 1,\tilde k \neq k}^K {\Bigl(1-I_n^{(\tilde k)}\Bigr)H_{m,n}^{(k,\tilde k)}}  + {P_N}}\Bigr)$, i.e., $\Gamma_{m,n}^{(k)}$ will increase by at least $\Delta$.


\revNN{
\section{Proof of Lemma~\ref{lemma:LP_relaxation}}
\label{appendix_lemma1}
We proceed by proving that ${\cal C} \subseteq {\cal C}^\prime$ and ${\cal C}^\prime \subseteq {\cal C}$. To prove that ${\cal C} \subseteq {\cal C}^\prime$, we assume that $y^{(k,\tilde k)}_{m,n} \in {\cal C}$ and deduce that $y^{(k,\tilde k)}_{m,n} \in {\cal C}^\prime$. Since $\sum_{\tilde k \in \tilde{\cal K}^{(k)}} y^{(k,\tilde k)}_{m,n} \leq 1$ and $y^{(k,\tilde k)}_{m,n} \leq x^{(k)}_{m,n}$, then $\sum_{\tilde k \in \tilde{\cal K}^{(k)}} y^{(k,\tilde k)}_{m,n} \leq x^{(k)}_{m,n}$.
Moreover, since $y^{(k,\tilde k)}_{m,n} \leq x^{(k)}_{m,n},$ then $\sum_{m=1}^{M^{(k)}} y^{(k,\tilde k)}_{m,n} \leq \sum_{m=1}^{M^{(k)}} x^{(k)}_{m,n} \leq 1$. Since $\sum_{m=1}^{M^{(k)}} y^{(k,\tilde k)}_{m,n} \leq 1$, $y^{(k,\tilde k)}_{m,n} \leq I^{(\tilde k)}_{n},$ and $y^{(k,\tilde k)}_{m,n} \in \{0,1\},$ then $\sum_{m=1}^{M^{(k)}} y^{(k,\tilde k)}_{m,n} \leq I^{(\tilde k)}_{n}$. Consequently, $y^{(k,\tilde k)}_{m,n} \in {\cal C}^\prime,$ which means that ${\cal C} \subseteq {\cal C}^\prime$.

Similarly, to show that ${\cal C}^\prime \subseteq {\cal C}$, we assume that $y^{(k,\tilde k)}_{m,n} \in {\cal C}^\prime$ and deduce that $y^{(k,\tilde k)}_{m,n} \in {\cal C}$. Since $\sum_{\tilde k \in \tilde{\cal K}^{(k)}} y^{(k,\tilde k)}_{m,n} \leq x^{(k)}_{m,n}$ and $x^{(k)}_{m,n}, y^{(k,\tilde k)}_{m,n} \in \{0,1\}$, then
$y^{(k,\tilde k)}_{m,n} \leq x^{(k)}_{m,n}$. Moreover, since $\sum_{m=1}^{M^{(k)}} y^{(k,\tilde k)}_{m,n} \leq I^{(\tilde k)}_{n}$ and $I^{(k)}_{n}, y^{(k,\tilde k)}_{m,n} \in \{0,1\}$, then $\sum_{m=1}^{M^{(k)}} y^{(k,\tilde k)}_{m,n} \leq 1$ which means that $y^{(k,\tilde k)}_{m,n} \leq I^{(\tilde k)}_{n}$. Hence, $y^{(k,\tilde k)}_{m,n} \in {\cal C}$ which means that ${\cal C}^\prime \subseteq {\cal C}$.

}

\revN{
\section{Proof of Proposition~\ref{lemma:LP_relaxation}}
\label{proof_lemma}

We start by categorizing the constraints of the relaxed version of~\eqref{ICICILP} into two groups: non-bounding constraints, which are given by~\eqref{ICICILP:Constraint1}, \eqref{ICICILP:Constraint2}, and \eqref{ICICILP:Constraint3}, and bounding constraints, which are given by~\eqref{LP_variables_bounds} (relaxed version of \eqref{ICICILP:Constraint4}). It is easy to see that the number of  variables that assume optimal binary values is equal to the number of bounding inequality constraints that are active at the optimal, i.e., inequalities that are satisfied with equalities at the optimal.

We now note that the optimal solution of a linear program lies on one or more vertices. For a linear program with $N_{var}$ variables, a vertex is characterized by $N_{var}$ equalities or active inequalities. Let $N^{\star}_{non-bounding}$ denote the number of non-bounding equality or active inequality constraints at the optimum, and let $N^{\star}_{bounding}$ denote the number of  active bounding inequality constraints at the optimum. Hence, we can write the following:
\begin{equation}
\begin{array}{l l l}
& N_{var}&= N^{\star}_{bounding}+N^{\star}_{non-bounding} \\
\Rightarrow &N^{\star}_{bounding}&= N_{var}-N^{\star}_{non-bounding}\\
\Rightarrow &N^{\star}_{bounding} &\geq  N_{var}-\sup(N^{\star}_{non-bounding})\\
& &= (\tilde{K}+1) \sum_{k=1}^K M^{(k)} +K \\
&&\ \ -(K+\sum_{k=1}^K (M^{(k)}+\tilde{K}))\\
&&= \sum_{k=1}^K \tilde{K}(M^{(k)}-1),
\end{array}
\end{equation}
where the supremum is found by assuming that all non-bounding constraints are active. Thus, the percentage of optimal variables of the relaxed version of~\eqref{ICICILP} that assume binary values can be bounded as
\begin{equation}
\begin{array}{l l}
P_\textrm{binary} &\geq \frac{\tilde{K}\sum_{k=1}^K (M^{(k)}-1)}{(\tilde{K}+1)\sum_{k=1}^K  M^{(k)} +K} \cdot 100\%\\  &=\frac{\tilde{K}(\bar{M}-1)}{(\tilde{K}+1)\bar{M}+1} \cdot 100\%,
\end{array}
\end{equation}
where $\bar{M}$ is the average number of UTs per sector, i.e., $\bar{M}=\frac{1}{K} \sum_{k=1}^{K} M^{(k)}$.
}
\begin{IEEEbiography}{Akram Bin Sediq} (S'04)
received the B. Sc. degree in electrical engineering (\textit{summa cum laude}) in 2006 from the American University of Sharjah, United Arab Emirates, and the M.A.Sc. degree in electrical engineering and the Ph.D. degree in electrical and computer engineering from Carleton University, Ottawa, Canada, in 2008 and 2013, respectively. In Winter 2012, Dr. Bin Sediq was a course instructor at the Department of Systems and Computer Engineering at Carleton University.

During 2013–-2014, Dr. Bin Sediq was a system engineer at BLiNQ Networks, Ottawa, Canada, where he was responsible for developing novel algorithms for non-line-of-sight small-cell backhaul networks, including, dual-carrier joint scheduling, interference avoidance, beam selection, and IEEE 1588 clock synchronization algorithms. Dr. Bin Sediq has recently joined the R\&D Group of Ericsson Canada, Ottawa, Canada.

His research  interests include inter-cell interference coordination, radio resource allocations, heterogenous networks, constellation design, cooperative communications, clock synchronization, and applications of optimization in wireless networks.

Dr. Bin Sediq is a recipient of the IEEE Antennas and Propagation Society Undergraduate Scholarship in 2004, the IEEE Microwave Theory and Techniques Society Undergraduate/Pre-Graduate Scholarship in 2005, the third prize in IEEE IAS Myron Zucker Undergraduate Student Design Contest in 2005, the President Cup for graduating with the highest GPA at the Bachelor's level in 2006, the Ontario Graduate Scholarship for international students for three years in a row during 2007--2010, two Senate Medals for outstanding academic achievement from Carleton University at the Masters and Ph.D. levels in 2008 and 2013, respectively, the EDC Teaching Assistant Outstanding Award at Carleton University in 2013, and
NSERC Industrial R\&D Fellowship~(IRDF), 2013--2015.
\end{IEEEbiography}

\begin{IEEEbiography}{Rainer Schoenen}(SM'2013) received his German Diplom-Ingenieur and Ph.D. degrees from RWTH Aachen University, in Electrical Engineering in 1995 and 2000, respectively. His research interests include stochastic Petri nets and queuing systems, ATM, TCP/IP, switching, flow control, QoS, tariffs, Userin-the-loop (UIL), wireless resource and packet scheduling and the MAC layer of 4+5G systems. His PhD thesis was ``System Components for Broadband Universal Networks with QoS Guarantee.'' with the ISS group of Prof. Heinrich Meyr at RWTH Aachen University, Germany, from 1995 to 2000. He started working self-employed in 2000. Dr. Schoenen was a senior researcher at the Communication Networks (ComNets) Research Group, RWTH Aachen with Professor Walke from 2005 to 2009, working on computer networks, queuing theory, Petri nets, LTE-Advanced, FDD relaying, scheduling, OSI layer 2 (MAC) and IMT-Advanced Evaluation within WINNER+. From 2010 to 2014 he worked at the Department of Systems and Computer Engineering, Carleton University, Ottawa, Canada, as a project manager supporting Professor Halim Yanikomeroglu. Since 2015 he is a professor at HAW Hamburg, Germany.
\end{IEEEbiography}

\begin{IEEEbiography}{Halim Yanikomeroglu}(S'96--M'98--SM'12) was born in Giresun, Turkey, in 1968. He received the B.Sc. degree in electrical and electronics engineering from the Middle East Technical University, Ankara, Turkey, in 1990, and the M.A.Sc. degree in electrical engineering (now ECE) and the Ph.D. degree in electrical and computer engineering from the University of Toronto, Canada, in 1992 and 1998, respectively.

During 1993–-1994, he was with the R\&D Group of Marconi Kominikasyon A.S., Ankara, Turkey. Since 1998 he has been with the Department of Systems and Computer Engineering at Carleton University, Ottawa, Canada, where he is now a Full Professor. His research interests cover many aspects of wireless technologies with a special emphasis on cellular networks. He coauthored over 60 IEEE journal papers, and has given a high number of tutorials and invited talks on wireless technologies in the leading international conferences. In recent years, his research has been funded by Huawei, Blackberry, Samsung, Communications Research Centre of Canada (CRC), Telus, and Nortel. This collaborative research resulted in about 20 patents (granted and applied). Dr. Yanikomeroglu has been involved in the organization of the IEEE Wireless Communications and Networking Conference (WCNC) from its inception, including serving as Steering Committee Member as well as the Technical Program Chair or Co-Chair of WCNC 2004 (Atlanta), WCNC 2008 (Las Vegas), and WCNC 2014 (Istanbul). He was the General Co-Chair of the IEEE Vehicular Technology Conference Fall 2010 held in Ottawa. He has served in the editorial boards of the IEEE TRANSACTIONS ON COMMUNICATIONS, IEEE TRANSACTIONS ON WIRELESS COMMUNICATIONS, and IEEE COMMUNICATIONS SURVEYS \& TUTORIALS. He was the Chair of the IEEE’s Technical Committee on Personal Communications (now called Wireless Technical Committee). He is a Distinguished Lecturer for the IEEE Vehicular Technology Society since 2010.

Dr. Yanikomeroglu is a recipient of the IEEE Ottawa Section Outstanding Educator Award in 2014, Carleton University Faculty Graduate Mentoring Award in 2010, the Carleton University Graduate Students Association Excellence Award in Graduate Teaching in 2010, and the Carleton University Research Achievement Award in 2009. Dr. Yanikomeroglu spent the 2011–2012 academic year at TOBB University of Economics and Technology, Ankara, Turkey, as a Visiting Professor. He is a registered Professional Engineer in the province of Ontario, Canada.
\end{IEEEbiography}

\begin{IEEEbiography}{Gamini Senarath} received the B.Sc. degree in electrical and electronics engineering from the Moratuwa University, Sri Lanka, in 1980, the Master of Electronics Engineering degree from the Eindhoven (Phillips)  International Institute, the Netherlands, in 1986 and the Ph.D. degree in telecommunications engineering from the University of Melbourne, Australia, in 1996, respectively.

During 1980--1990, he was with the public telecommunication provider of the Government of Sri Lanka, as a Radio Transmission Engineer and a Regional Telecommunication Manager. During 1996 to 2009 he was with the Advanced Wireless Technology Labs, Nortel Networks, Ottawa, Canada, where he was involved in many next generation wireless system research projects and contributed to various 3G and 4G wireless standard development activities. The main focus was on propagation, cell planning, relaying and MAC/RRM areas. Since 2009, he has been with 5G Research Lab, Huawei Technologies Canada. His current research work involves software defined networks, virtual network embedding, power and the resource management of 3rd party infra-structure provided by multiple operators for the 5G wireless systems. His work resulted in more than 50 granted patents in wireless communication systems. Dr. Senarath is also the project manager for the Huawei's external collaboration project with Carleton University, Telus and Communications Research Centre of Canada (CRC).
\end{IEEEbiography}


\begin{thebibliography}{10}
\providecommand{\url}[1]{#1}
\csname url@samestyle\endcsname
\providecommand{\newblock}{\relax}
\providecommand{\bibinfo}[2]{#2}
\providecommand{\BIBentrySTDinterwordspacing}{\spaceskip=0pt\relax}
\providecommand{\BIBentryALTinterwordstretchfactor}{4}
\providecommand{\BIBentryALTinterwordspacing}{\spaceskip=\fontdimen2\font plus
\BIBentryALTinterwordstretchfactor\fontdimen3\font minus
  \fontdimen4\font\relax}
\providecommand{\BIBforeignlanguage}[2]{{%
\expandafter\ifx\csname l@#1\endcsname\relax
\typeout{** WARNING: IEEEtran.bst: No hyphenation pattern has been}%
\typeout{** loaded for the language `#1'. Using the pattern for}%
\typeout{** the default language instead.}%
\else
\language=\csname l@#1\endcsname
\fi
#2}}
\providecommand{\BIBdecl}{\relax}
\BIBdecl

\bibitem{ABS_PIMRC2011}
A.~{Bin Sediq}, R.~Schoenen, H.~Yanikomeroglu, G.~Senarath, and Z.~Chao, ``A
  novel distributed inter-cell interference coordination scheme based on
  projected subgradient and network flow optimizations,'' in \emph{{Proc. IEEE
  Int. Symp. on Pers., Indoor and Mobile Radio Commun.~(PIMRC)}}, Toronto,
  Canada, Sep. 2011.

\bibitem{Rappaport}
T.~S. Rappaport, \emph{Wireless Communications: Principles and Practice}.\hskip
  1em plus 0.5em minus 0.4em\relax Prentice-Hall, 1996.

\bibitem{Rahman2010}
M.~Rahman and H.~Yanikomeroglu, ``Enhancing cell-edge performance: A downlink
  dynamic interference avoidance scheme with inter-cell coordination,''
  \emph{{IEEE Trans.\ Wireless Commun.}}, vol.~9, no.~4, pp. 1414--1425, Apr.
  2010.

\bibitem{SFR2005}
{3GPP Project Doc. R1-050 507}, ``Soft frequency reuse scheme for {UTRAN
  LTE},'' Tech. Rep., May 2005.

\bibitem{Sternad2003}
M.~Sternad, T.~Ottosson, A.~Ahlen, and A.~Svensson, ``Attaining both coverage
  and high spectral efficiency with adaptive {OFDM} downlinks,'' in
  \emph{{Proc. IEEE Veh. Tech. Conf.~(VTC-Fall)}}, Orlando, Florida, Oct. 2003.

\bibitem{ComparingPFRSFR}
Y.~Xiang, J.~Luo, and C.~Hartmann, ``Inter-cell interference mitigation through
  flexible resource reuse in {OFDMA} based communication,'' in \emph{Proc. 13th
  European Wireless Conf.}, Paris, France, Apr. 2007.

\bibitem{VTC2007Comparison}
A.~Simonsson, ``Frequency reuse and intercell interference co-ordination in
  {E-UTRA},'' in \emph{{Proc. IEEE Veh. Tech. Conf.~(VTC-Spring)}}, Dublin,
  Ireland, Apr. 2007.

\bibitem{AdaptiveFFR}
{3GPP Project Doc. R1-072 762}, ``Further discussion on adaptive fractional
  frequency reuse,'' Tech. Rep., Jun. 2007.

\bibitem{Hussain2009}
S.~H. Ali and V.~C.~M. Leung, ``Dynamic frequency allocation in fractional
  frequency reused {OFDMA} networks,'' \emph{{IEEE Trans.\ Wireless Commun.}},
  vol.~8, no.~12, pp. 4286--4295, Aug. 2009.

\bibitem{Zhang2008}
X.~Zhang, C.~He, L.~Jiang, and J.~Xu, ``Inter-cell interference coordination
  based on softer frequency reuse in {OFDMA} cellular systems,'' in \emph{Proc.
  Int. Conf. on Neural Net. and Signal Process.~(ICNNSP)}, Zhenjiang, China,
  Jun. 2008.

\bibitem{Mao2008}
X.~Mao, A.~Maaref, and K.~H. Teo, ``Adaptive soft frequency reuse for
  inter-cell interference coordination in {SC-FDMA} based {3GPP LTE} uplinks,''
  in \emph{{Proc. IEEE Glob. Commun. Conf.~(Globecom)}}, New Orleas, LA, Dec.
  2008.

\bibitem{Stoylar2009}
A.~Stolyar and H.~Viswanathan, ``Self-organizing dynamic fractional frequency
  reuse for best-effort traffic through distributed inter-cell coordination,''
  in \emph{{Proc.\ IEEE Int.\ Conf.\ Comp.\ Commun.~(INFOCOM)}}, Rio de
  Janeiro, Brazil, Apr. 2009.

\bibitem{DBNRS_PIMRC10}
D.~B\"{u}ltmann, T.~Andre, and R.~Schoenen, ``{Analysis of 3GPP LTE-Advanced
  cell spectral efficiency},'' in \emph{{Proc. IEEE Int. Symp. on Pers., Indoor
  and Mobile Radio Commun.~(PIMRC)}}, Istanbul, Turkey, Sep. 2010.

\bibitem{Li2006}
G.~Li and H.~Liu, ``Downlink radio resource allocation for multi-cell {OFDMA}
  system,'' \emph{{IEEE Trans.\ Wireless Commun.}}, vol.~5, no.~12, pp. 3451
  --3459, Dec. 2006.

\bibitem{Chang2009}
R.~Y. Chang, Z.~Tao, J.~Zhang, and C.~C.~J. Kuo, ``Multicell {OFDMA} downlink
  resource allocation using a graphic framework,'' \emph{{IEEE Trans. Veh.
  Technol.}}, vol.~58, no.~12, pp. 3494--3507, Sep. 2009.

\bibitem{Ellenbeck2008}
J.~Ellenbeck, C.~Hartmann, and L.~Berlemann, ``Decentralized inter-cell
  interference coordination by autonomous spectral reuse decisions,'' in
  \emph{Proc. 14th European Wireless Conf.}, Prague, Czech Republic, Jun. 2008.

\bibitem{TomLuo2008}
Z.-Q.~T. Luo and S.~Zhang, ``Dynamic spectrum management: Complexity and
  duality,'' \emph{{IEEE J.\ Select.\ Topics Signal Processing}}, vol.~2,
  no.~1, pp. 57--73, Feb. 2008.

\bibitem{Network_flows}
R.~Ahuja, T.~Magnanti, and J.~Orlin, \emph{{Network Flows: Theory, Algorithms,
  and Applications}}.\hskip 1em plus 0.5em minus 0.4em\relax Prentice Hall,
  1993.

\bibitem{Tao2013_MCNF}
M.~Tao and Y.~Liu, ``A network flow approach to throughput maximization in
  cooperative {OFDMA} networks,'' \emph{{IEEE Trans.\ Wireless Commun.}},
  vol.~12, no.~3, pp. 1138--1148, Mar. 2013.

\bibitem{Zaki2011_MCNF}
A.~N. Zaki and A.~O. Fapojuwo, ``Optimal and efficient graph-based resource
  allocation algorithms for multiservice frame-based {OFDMA} networks,''
  \emph{{IEEE Trans. Mobile Comput.}}, vol.~10, no.~8, pp. 1175--1186, Aug.
  2011.

\bibitem{IMTAdvanced}
\BIBentryALTinterwordspacing
{ITU, Report~ITU-R~M.2135-1}, ``Guidelines for evaluation of radio interface
  technologies for {IMT-Advanced},'' ITU, Tech. Rep., Dec. 2009. [Online].
  Available: \url{http://www.itu.int/pub/R-REP-M.2135-1-2009.}
\BIBentrySTDinterwordspacing

\bibitem{Kushner}
H.~Kushner and P.~Whiting, ``Convergence of proportional-fair sharing
  algorithms under general conditions,'' \emph{{IEEE Trans.\ Wireless
  Commun.}}, vol.~3, no.~4, pp. 1250--1259, Jul. 2004.

\bibitem{StoylarGradient2005}
A.~L. Stolyar, ``On the asymptotic optimality of the gradient scheduling
  algorithm for multiuser throughput allocation,'' \emph{{Oper. Res.}},
  vol.~53, no.~1, pp. 12--25, Jan. 2005.

\bibitem{SongPart2}
G.~Song and Y.~Li, ``Cross-layer optimization for {OFDM} wireless
  networks---part {II}: Algorithm development,'' \emph{{IEEE Trans.\ Wireless
  Commun.}}, vol.~4, no.~2, pp. 625--634, Mar. 2005.

\bibitem{GeneralizedPF}
J.~Mo and J.~Walrand, ``Fair end-to-end window-based congestion control,''
  \emph{{IEEE/ACM Trans. Netw.}}, vol.~8, no.~5, pp. 556--567, Oct. 2000.

\bibitem{ChiangFairness}
T.~Lan, D.~Kao, M.~Chiang, and A.~Sabharwal, ``An axiomatic theory of fairness
  in network resource allocation,'' in \emph{{Proc.\ IEEE Int.\ Conf.\ Comp.\
  Commun.~(INFOCOM)}}, San Diego, California, Mar. 2010.

\bibitem{KellyPF}
F.~P. Kelly, A.~K. Maulloo, and D.~K.~H. Tan, ``Rate control for communication
  networks: Shadow prices, proportional fairness and stability,'' \emph{{J.
  Oper. Res. Soc.}}, vol.~49, no.~3, pp. 237--252, 1998.

\bibitem{OpportunisticBeamforming}
P.~Viswanath, D.~Tse, and R.~Laroia, ``Opportunistic beamforming using dumb
  antennas,'' \emph{{IEEE Trans.\ Inf.\ Theory}}, vol.~48, no.~6, pp. 1277
  --1294, Jun. 2002.

\bibitem{JTWC2013}
A.~{Bin~Sediq}, R.~H. Gohary, R.~Schoenen, and H.~Yanikomeroglu, ``Optimal
  tradeoff between sum-rate efficiency and {J}ain's fairness index in resource
  allocation,'' \emph{{IEEE Trans.\ Wireless Commun.}}, vol.~12, no.~7, pp.
  3496--3509, Jul. 2013.

\bibitem{EffeciencyFairness3}
H.~T. Cheng and W.~Zhuang, ``An optimization framework for balancing throughput
  and fairness in wireless networks with {QoS} support,'' \emph{{IEEE Trans.\
  Wireless Commun.}}, vol.~7, no.~2, pp. 584--593, Feb. 2008.

\bibitem{Stolyar_PF}
M.~Andrews, L.~Qian, and A.~Stolyar, ``Optimal utility based multi-user
  throughput allocation subject to throughput constraints,'' in \emph{{Proc.\
  IEEE Int.\ Conf.\ Comp.\ Commun.~(INFOCOM)}}, Miami, Florida, Mar. 2005.

\bibitem{Bertsekas_book}
D.~P. Bertsekas, \emph{Nonlinear Programming}, 2nd~ed.\hskip 1em plus 0.5em
  minus 0.4em\relax Athena Scientific, 2008.

\bibitem{PrimalDecomposition}
S.~Boyd, L.~Xiao, A.~Mutapcic, and J.~Mattingley, ``Notes on decomposition
  methods,'' Lecture notes for EE364B, Stanford University, Oct. 2003.

\bibitem{SubgradientMethod}
S.~Boyd and A.~Mutapcic, ``Subgradient methods,'' Lecture notes for EE364b,
  Stanford University, Oct. 2003.

\bibitem{Subgradients}
S.~Boyd, ``Subgradients,'' Lecture notes for EE364b, Stanford University, Apr.
  2011.

\bibitem{IMTChannelModel}
\BIBentryALTinterwordspacing
J.~Nystrom, ``Software implementation of {IMT.Eval} channel model,''
  CELTIC/CP5-026 Project WINNER+ Doc. 5D/478-E, Tech. Rep., Jul. 2009.
  [Online]. Available: \url{http://www.itu.int/oth/R0A06000022/en.}
\BIBentrySTDinterwordspacing

\bibitem{RS_WICOM10}
R.~Schoenen and C.~Teijeiro, ``{System level performance evaluation of LTE with
  MIMO and relays in reuse-1 IMT-Advanced scenarios},'' in \emph{Proc. IEEE
  WiCom}, Chengdu, China, Sep. 2010.

\bibitem{WinnerCalibration}
\BIBentryALTinterwordspacing
``Calibration for {IMT-Advanced} evaluations,'' {CELTIC/CP5-026 Project
  WINNER+}, Tech. Rep., May 2010. [Online]. Available:
  \url{http://projects.celtic-initiative.org/winner+/WINNER+ Evaluation
  Group.html.}
\BIBentrySTDinterwordspacing

\bibitem{VTC2008_OptimumFFR}
M.~Assaad, ``Optimal fractional frequency reuse ({FFR}) in multicellular
  {OFDMA} system,'' in \emph{{Proc. IEEE Veh. Tech. Conf.~(VTC-Fall)}}, Marina
  Bay, Singapore, Sep. 2008.

\bibitem{RS2011PIMRC}
R.~Schoenen, A.~{Bin Sediq}, H.~Yanikomeroglu, G.~Senarath, and Z.~Chao,
  ``Fairness analysis in cellular networks using stochastic {Petri} nets,'' in
  \emph{{Proc. IEEE Int. Symp. on Pers., Indoor and Mobile Radio
  Commun.~(PIMRC)}}, Toronto, Canada, Sep. 2011.

\bibitem{decomposition_TSP2014}
G.~Scutari, F.~Facchinei, P.~Song, D.~P. Palomar, and J.-S. Pang,
  ``Decomposition by partial linearization: Parallel optimization of
  multi-agent systems,'' \emph{{IEEE Trans.\ Signal Process.}}, vol.~62, no.~3,
  pp. 641--656, Feb. 2014.

\end{thebibliography}
\end{document}